\documentclass[twocolumn,showpacs,aps]{revtex4}
\usepackage{lipsum}
\usepackage{color}
\usepackage{graphicx}
\usepackage{pstricks}
\usepackage{dcolumn}
\usepackage{bm}
\usepackage{epsf}
\usepackage{epsfig}
\usepackage{amsmath,amsfonts,amsthm, amssymb,latexsym}
\usepackage{enumerate}

\newcommand{\beq}{\begin{equation}}
\newcommand{\eeq}{\end{equation}}
\newcommand{\bcn}{\begin{center}}
\newcommand{\ecn}{\end{center}}

\newcommand{\lsim}{\lower0.5ex\hbox{$\; \buildrel < \over \sim \;$}}

\begin{document}

\title{Many-body forces in the equation of state of hyperonic matter}

\author{R.O. Gomes}
 \email{rosana.gomes@ufrgs.br}
\affiliation{Instituto de F\'isica, Universidade Federal do Rio Grande do Sul, Porto Alegre, RS 91501-970, Brazil}
\affiliation{Frankfurt Institute for Advanced Studies, Goethe University, D-60438 Frankfurt am Main, Germany}
        
 \author{V.A. Dexheimer}
\affiliation{Department of Physics, Kent State University, Kent, OH 44242, USA}


\author{S. Schramm}
\affiliation{Frankfurt Institute for Advanced Studies, Goethe University, D-60438 Frankfurt am Main, Germany}

\author{C.A.Z. Vasconcellos}
\affiliation{Instituto de F\'isica, Universidade Federal do Rio Grande do Sul, Porto Alegre, RS 91501-970, Brazil}
\affiliation{ICRANet, Piazza della Repubblica 10, 65122 Pescara, Italy}

\date{\today}

\begin{abstract}
In this work we introduce an extended version of the formalism proposed originally by Taurines et al. that considers 
the effects of many-body forces simulated by non-linear self-couplings and meson-meson interaction contributions.
In this extended version of the model, we assume that matter is at zero temperature, charge neutral and in beta-equilibrium,
considering that the baryon octet 
interacts by the exchange of scalar-isoscalar ($\sigma$,$\,\sigma^*$), vector-isoscalar ($\omega$,$\,\phi$), vector-isovector ($\varrho$)
and scalar-isovector ($\delta$) meson fields. 
Using nuclear matter properties, we constrain the parameters of the model
that describe the intensity of the indirectly density dependent baryon-meson couplings to a small range of possible values.
We then investigate asymmetric hyperonic matter properties. 
We report that the formalism developed in this work 
is in agreement with experimental data and also allows for the existence of massive hyperon stars (with more than $2M_{\odot}$) 
with small radii, compatible with astrophysical observations.  

\end{abstract}

\maketitle
\section{Introduction} \label{sec:Int}

Neutron stars are formed when stars with initial masses of about $8-20\,M_{\odot}$ collapse in supernovae events. 
During this process, the outer layers of the star exert pressure upon the central region, increasing its density and,
when the system reaches the drip line density, the nucleons inside the nuclei are released. 
After the mass shock wave bounces from the core, the outer parts of the star are released in an explosive reaction, 
leaving a dense core behind.
Since the scale of Fermi energy present in nuclear processes in neutron stars is much higher than the thermal energy, 
these objects provide a unique scenario to study the behavior of nuclear matter at extreme conditions.

Experimental nuclear data together with neutron star observational measurements
can and should be used to restrict the parameters of the models used to describe 
the equation of state (EoS) of neutron stars, not only at low
but also at high densities.
For example, ref. \cite{SchaffnerBielich:2008kb} showed that hypernuclear data impose
contraints in the composition of hyperon stars (see also references cited in \cite{SchaffnerBielich:2008kb}).
More specifically, observations of $~2M_{\odot}$ neutron stars \cite{Demorest2010,Antoniadis2013} 
constrain the EoS of neutron stars, in particular its stiffness in the high density regime.
A theoretical study that calculates the radii of $1.4M_{\odot}$ neutron stars
also discusses the role played by matter above saturation density on neutron star properties \cite{Fortin:2014mya}. 
Such compact stars are considered \emph{canonical stars} because the most accurate mass measurement 
from the binary pulsar system PSR 1913+16, predicts masses of $1.3867\pm0.0002M_{\odot}$ and $1.4414\pm0.0002M_{\odot}$ 
\cite{Weisberg:2004hi,Lattimer:2004sa}.

In the last decades, the determination of the EoS of nuclear matter at high densities has become one 
of the main goals of nuclear astrophysics.
Although the fundamental physics has to be described in terms of quarks and gluons, 
at scales of energy where only the baryon degrees of freedom are relevant, the residuum of 
the strong interaction between quarks can be described by effective hadronic models. 
In a particular class of models, denominated 
\emph{relativistic mean field (RMF)} models, the nuclear interaction is usually described by the exchange of scalar-isoscalar and 
vector-isoscalar mesons with minimal Yukawa coupling. 
These terms represent the attractive and repulsive components of the nuclear force 
in the long and short range regimes, respectively.

The first RMF model for nuclear matter was proposed in 1974 by Walecka et al. \cite{Walecka1986}, 
introducing the description of nuclear matter in the exchange  of scalar and vector mesons. 
This model, however, was not able to reproduce correctly the 
compressibility modulus of nuclear matter and the nucleon effective mass at saturation.
As a consequence, several extensions of the model were developed in order to overcome the problem.

Boguta and Bodmer proposed additional terms in the scalar sector by taking into account the 
non-linear third and fourth orders of the $\sigma$ meson self-interaction terms.
With this, they could simulate field-dependent correlations
in the {\it in-medium} nucleon-nucleon interaction \cite{Boguta:1977xi}.
In later works, this same feature was extended to the $\omega$ meson \cite{Sugahara:1993wz, Toki1995} 
and the case with crossed scalar-vector interaction terms was also investigated \cite{ToddRutel:2005zz,Kumar:2006ij}.
The meson-baryon coupling was also modified in other models, as an alternative to the minimal coupling 
proposed by Walecka. In particular, we mention the ZM models with a derivative coupling \cite{Zimanyi:1990np}
and the density dependent meson-baryon coupling model proposed in ref. \cite{Typel:1999yq}. 

Since the characteristic time scale for the population of neutron stars is long compared to the typical 
weak-interaction time scale, strangeness is not conserved in their interiors, allowing for the 
appearance of hyperonic degrees of freedom.
The topic of hyperons in neutron stars has been extensively discussed in the literature.
In particular, the discussion relating new degrees of freedom to the softening of the EoS and, consequently, 
to the lowering of neutron stars maximum masses received renewed attention 
\cite{Sahakian1963,Glendenning:1982nc,Glendenning:1984jr,Knorren:1995ds,Hanauske:1999ga,SchaffnerBielich:2002ki} 
due to the observations of massive neutron stars.
However, as first shown in \cite{Schaffner:1995th}, this problem is avoided by the addition
of the pair of strange mesons $\sigma^*$ and $\phi$ (associated, respectively, to $f_0(975\,\mathrm{MeV})$ and $\phi(1020\,\mathrm{MeV})$ \cite{Beringer:1900zz}),
where the latter one introduces a new repulsion to the interaction of hyperons. 
Although additional uncertainties concerning the hyperonic coupling are added to the formalism \cite{Fortin:2014mya},
in the presence of these additional mesons, several authors succeeded in describing massive hyperon stars using different models 
\cite{Dexheimer:2008ax, Bednarek:2011gd, Bonanno:2011ch, Weissenborn:2011ut, Weissenborn:2011kb,  
Lopes:2013cpa, Lastowiecki:2011hh, Bhowmick:2014pma, Banik:2014qja, Gusakov:2014ota, vanDalen:2014mqa, Yamamoto:2014jga}.

The formulation used in this work is based on the model proposed by Taurines et al 
in 2001 \cite{Taurines:2000zb} as an attempt to unify the Walecka and ZM models in a general fashion.
This formalism considers a parametric derivative coupling that simulates the many-body forces by 
including non-linear self-interaction and meson-meson interaction terms for the scalar mesons in the Lagrangian density of the theory.
The meson-baryon couplings are implicitly density dependent and parametrized by a constant.
The parametric coupling formalism reproduces successfully nuclear matter properties and has been
applied to investigate different topics related to nuclear physics, such as the nuclear matter compressibility \cite{Dexheimer:2007mt},
hadron-quark phase transitions \cite{Burigo:2010zz}, kaon condensation \cite{Razeira:2011zz}, symmetry energy \cite{Marranghello:2010wu}
and effects of magnetic fields on neutron stars \cite{Gomes:2014ova,Gomes:2014dka}.

In this work, we present an extended version of the parametric model including the entire set of mesons
relevant to the scale of energy and mean field approximation: scalar-isoscalar ($\sigma$,$\,\sigma^*$), 
vector-isoscalar ($\omega$,$\,\phi$), vector-isovector ($\varrho$) and scalar-isovector ($\delta$) meson fields.
The $\varrho$ and $\delta$ mesons are very important for the
description of neutron stars, as these are highly isopin asymmetric objects \cite{Kubis:1997ew,Liu:2001iz,Menezes:2004vr}. 
Also, the strange mesons $\phi$ and $\sigma^*$, which are often disregarded, are important for the description of 
hyperon-hyperon interaction, and hence have an important impact on the description of neutrons stars.

The model presented in this paper is constrained by finding the parameterization that best describes symmetric nuclear matter properties (effective nucleon mass and the 
compressibility modulus at saturation density) and asymmetric matter properties (symmetry energy and its slope), for fixed values of saturation density and 
binding energy. 
We determine the hyperon-meson coupling constants for the vector mesons by the SU(6) symmetry relations 
and the scalar meson coupling by the fitting of hyperon potentials to experimental data.
We analyze how nuclear matter properties, hyperon-nucleon and hyperon-hyperon interactions affect the predictions of the neutron star properties. 
Also, we identify the effects of the new mesonic degrees of freedom on these properties.
 
The paper is organized as follows: In Sec. II we present the many-body formalism introduced in the Lagrangian density and analyze
its effects on the EoS and chemical equilibrium for the model; the nuclear matter properties at saturation density are calculated in Sec. III,
when we set the parameters of the model; section IV is dedicated to the study of asymmetric matter; 
we discuss the astrophysical applications of the model in the description of hyperon stars
concerning the meson fields and hyperon interactions in Sec. V and, finally, we present our conclusions in Sec. VI.

 
\section{Many-body coupling relativistic field formalism} \label{Model}

\subsection{Formalism} 
As a conventional way of classifying and organizing interaction terms in effective field theory approaches, 
as well as to introduce a guideline for the strengths of the various couplings, we adopt the concept of naturalness.
Naturalness is related to effective interactions field theories that can be truncated within a phenomenological
domain of the theory.
Of course, there is no general proof of naturalness property, since no one knows how to derive the effective strong 
interaction Lagrangian density from QCD. 
Nevertheless, the validity of naturalness is supported by phenomenology (see e.g. \cite{Vasconcellos:2014qua}).

The formalism developed in ref. \cite{Taurines:2000zb} takes into account many-body contributions to the nuclear force 
by introducing the concept of naturalness and a parameterized derivative coupling for the mesons.
In this extended version of the formalism, we introduce the new mesonic degrees of freedom $\delta$, $\sigma^*$ and $\phi$.
The $\delta$ meson is introduced in order to better describe the properties of asymmetric matter, while 
the strange mesons ($\sigma^*$, $\phi$) have important impact on hyperon interactions.
The general Lagrangian density of the model is:

\small
\begin{equation}\begin{split}\label{lagrangiana_zen2}
\mathcal{L}&= \underset{b}{\sum}\overline{\psi}_{b}\left[\gamma_{\mu}\left(i\partial^{\mu}-g_{\omega b \xi}^{*}\omega^{\mu} -g_{\phi b \kappa}^{*}\phi^{\mu}
-\frac{1}{2}g_{\varrho b \eta}^{*}\mathbf{\boldsymbol{\textrm{\ensuremath{\tau}.\ensuremath{\varrho^{\mu}}}}}\right) \right.
\\& \left. - \left(1+\frac{g_{\sigma b}\sigma+g_{\sigma^* b}\sigma^*
+\frac{1}{2}g_{\delta b}\boldsymbol{\tau.\delta}}{\zeta m_{b}}\right)^{-\zeta} m_b\right]\psi_{b}
\\& +\left(\frac{1}{2}\partial_{\mu}\sigma\partial^{\mu}\sigma-m_{\sigma}^{2}\sigma^{2}\right)
+\left(\frac{1}{2}\partial_{\mu}\sigma^*\partial^{\mu}\sigma^*-m_{\sigma^*}^{2}\sigma^{* 2}\right)
\\&+\frac{1}{2}\left(-\frac{1}{2}\omega_{\mu\nu}\omega^{\mu\nu}+m_{\omega}^{2}\omega_{\mu}\omega^{\mu}\right)
+\frac{1}{2}\left(-\frac{1}{2}\phi_{\mu\nu}\phi^{\mu\nu}+m_{\phi}^{2}\phi_{\mu}\phi^{\mu}\right)
\\&+\frac{1}{2}\left(-\frac{1}{2}\boldsymbol{\varrho_{\mu\nu}.\varrho^{\mu\nu}}+m_{\varrho}^{2}\boldsymbol{\varrho_{\mu}.\varrho^{\mu}}\right)
+\left(\frac{1}{2}\partial_{\mu}\boldsymbol{\delta.}\partial^{\mu}\boldsymbol{\delta}-m_{\delta}^{2}\boldsymbol{\delta}^{2}\right)
\\& +\underset{l}{\sum}\overline{\psi}_{l}\gamma_{\mu}\left(i\partial^{\mu}-m_{l}\right)\psi_{l}.
\end{split}\end{equation}
\normalsize

The subscripts $b$ and $l$ label, respectively, the baryon octet ($n$, $p$, $\Lambda^0$, $\Sigma^-$, $\Sigma^0$, $\Sigma^+$, $\Xi^-$, $\Xi^0$)  
and lepton ($e^-$, $\mu^-$) degrees of freedom. 
The first and last terms represent the Dirac Lagrangian density for baryons and leptons, respectively. 
The other terms represent the Lagrangian densities of the mesons, where we assume a
Klein-Gordon Lagrangian density for the scalar $\sigma$, $\delta$ and $\sigma^*$ fields and a Proca Lagrangian density for the
vector $\omega$, $\varrho$ and $\phi$ fields. The meson-baryon coupling is introduced by the coupling constants present in the first term
of Equation \ref{lagrangiana_zen2}. 
The operators $\boldsymbol{\tau}=(\tau_1,\,\tau_2,\,\tau_3)$ denote the 
Pauli isospin matrices. 
We allow the system to be isospin asymmetric by coupling the $\delta$ and $\varrho$ 
fields to the isospin-dependent scalar ($\rho_s=\overline{\psi}\boldsymbol{\tau}\psi$) and vector densities ($\rho_b=\psi^{\dagger}\boldsymbol{\tau}\psi$).
The baryons, leptons and mesons properties are found in Tables (\ref{particles}) and (\ref{campos}).

\begin{table}[!ht]
\caption{Baryon and lepton properties. The rows indicate different particles and $I^3$, $q_b$, $q_e$ and $s$ stand
for the isopin projection in the $z$-direction, baryon charge, electric charge and strangeness, respectively.} 
\begin{center} \label{particles}
\begin{tabular}{|c|c|c|c|c|c|c|c|c|} 
\hline 
Particle & Mass $(\mathrm{MeV})$ & $I^{3}$ & $q_b$ & $q_e$ & $s$\tabularnewline
\hline
\hline
$p$ & $939.6$ & $1/2$ & $1$ & $+1$ & $0$ \tabularnewline
\hline
$n$ & $938.3$ & $-1/2$ & $1$ & $0$ & $0$ \tabularnewline
\hline
$\Lambda$ & $1116$ & $0$ & $1$ & $0$ & $-1$ \tabularnewline
\hline
$\Sigma^{+}$ & $1189$ & $+1$ & $1$ & $+1$ & $-1$\tabularnewline
\hline
$\Sigma^{0}$ & $1193$ & $0$ & $1$ & $0$ & $-1$ \tabularnewline
\hline
$\Sigma^{-}$ & $1197$ & $-1$ & $1$  & $-1$ & $-1$ \tabularnewline
\hline
$\Xi^{0}$ & $1315$ & $+1/2$ & $1$ & $0$ & $-2$ \tabularnewline
\hline
$\Xi^{-}$ & $1321$ & $-1/2$ & $1$  & $-1$ & $-2$ \tabularnewline
\hline
$e^{-}$ & $0.511$ & $0$ & $0$ & $-1$ & $0$ \tabularnewline
\hline
$\mu^{-}$ & $105.7$ & $0$ & $0$  & $-1$ & $0$ \tabularnewline
\hline\hline
\end{tabular}
\par\end{center}
\end{table}

\begin{table}[t]
\caption{Meson fields properties considered in the formalism.} 
\begin{center} \label{campos}
\begin{tabular}{|c|c|c|c|c|} 
\hline 
Meson & Particle & Classification & Coupling  & Mass \\
& & & constant & (MeV) \\ 
\hline \hline
$\sigma $ &  $\sigma $ & scalar-isoscalar & $g_{\sigma_b}$ & 550 \tabularnewline
$\mbox{\boldmath$\delta$}$  & $a_0$& scalar-isovector &$g_{\delta_b}$ &980\tabularnewline
$\omega_\mu $ & $\omega $& vector-isoscalar &$g_{\omega_b}$ &782 \tabularnewline
$\mbox{\boldmath$\varrho$}_{\mu}$  & $\rho$ & vector-isovector & $g_{\varrho_b}$&770 \tabularnewline
$\sigma^\ast $ & $f_0$ & scalar-isoscalar &$g_{\sigma^*_b}$ & 975 \tabularnewline
$\phi_\mu $ & $\phi $ & vector-isoscalar & $g_{\phi_b}$ & 1020 \tabularnewline
\hline\hline
\end{tabular}
\end{center}
\end{table}

The general definition of the meson-baryon couplings are:
\begin{equation}\label{coupling_zen}
g_{\omega b \xi}^{*}\equiv m_{\xi b}^{\star} g_{\omega b},
\quad g_{\varrho b \kappa}^{*}\equiv m_{\kappa b}^{\star} g_{\varrho b},
\quad g_{\phi b \eta}^{*}\equiv m_{\eta b}^{\star} g_{\phi b},\end{equation}
where the parametric coefficient $m_{\lambda b}^{\star}$ introduces the nonlinear contributions:
\begin{equation}\label{mstar_zen}
m_{\lambda b}^{\star}\equiv \left(1+\frac{g_{\sigma b}\sigma+g_{\sigma^* b}\sigma^*
+\frac{1}{2}g_{\delta b}\boldsymbol{\tau.\delta}}{\lambda\,m_{b}}\right)^{-\lambda},
\end{equation}
for $\lambda= \xi,\,\kappa,\,\eta,\,\zeta$.

The main motivation for this formalism is to introduce a parameterized derivative coupling that can be expanded in a series of
nonlinear couplings terms between the scalar $\sigma$, $\sigma^*$ and $\delta$ mesons. 
Each term of the expansion correponds to a medium effect contribution from many-body forces. 
Ultimately, the complete series expansion, controlled by the
$\lambda=\,\xi,\,\kappa,\,\eta,\,\zeta$ parameter, allows the description 
of genuine many-body forces, which are introduced as medium effects in the mean field approximation.

Each set of parameters generates different EoS's and population profiles 
and must be analyzed to cover the range of uncertainties of nuclear saturation properties.
We emphasize that, since the meson fields change their values with density, the scalar mesons 
coupling constants present a density dependence through the coupling of the fields.
Note, however, that such approach is thermodynamically consistent.

As a first approach, we consider the so-called \emph{scalar version} of the model, in which the nonlinear
contributions affect only the scalar mesons, i.e., $\xi=0,\,\kappa=0,\,\eta=0,\,\zeta\neq 0$.
See ref. \cite{Vasconcellos:2014qua} for a discussion of other possible paraterization choices.

The effective mass of baryons in this model reads:
\begin{equation}\begin{split}\label{meff_zen}
 m^*_{b \zeta}= & \, m_{\zeta b}^{\star} m_b \\
 \equiv& \left(1+\frac{g_{\sigma b}\sigma+g_{\sigma^* b}\sigma^* +\frac{1}{2}g_{\delta b}\boldsymbol{\tau.\delta}}{\zeta m_{b}}\right)^{-\zeta} m_b,
\end{split}\end{equation}
where one can see the influence of the nonlinear contributions of scalar mesons, controlled through the parameter $\zeta$.
Furthermore, as the effective mass of baryons depends on the $\zeta$-parameter, the chemical potential of
the interacting particles is also affected by many-body forces as:
\begin{equation}\label{mueff}
 \mu^*_{b}= \sqrt{k_{f_b}^2+(m_{b \zeta}^{*})^2} + g_{\omega b}\omega_0 + g_{\varrho b}I_{3 b} \varrho_{03} + g_{\phi b}\phi_0,
\end{equation}
where $k_{f_b}$ is the baryon Fermi momentum and $I_{3 b}$ is the baryon isospin projection in the $z$-direction.

The mean field equations, calculated from the Lagrangian density in the mean field approximation, read:

\begin{equation}\begin{split}\label{tcm}
\sigma_{0}&=\frac{1}{m_{\sigma}^{2}}\underset{b}{\sum}
\left[g_{\sigma b}\left(m_{\zeta b}^{\star}\right) \right.
\\ & \left. -\frac{g_{\sigma b}}{m_b}
\left(m_{\zeta b}^{\star}\right)^{\frac{\zeta+1}{\zeta}}
\left(g_{\sigma b}\sigma_{0} + g_{\delta b}\delta_{0}^{3}\,I^{3b}+g_{\sigma_{0}^* b}\sigma^*\right)\right]\rho_{s b}, \\
\omega_{0}&=\frac{1}{m_{\omega}^{2}}\underset{b}{\sum}g_{\omega b}\rho_{b}, \\
\varrho_{0}^{3}&=\frac{1}{m_{\varrho}^{2}}\underset{B}{\sum}g_{\varrho b}\, I^{3b}\,\rho_{b},\\
\delta_{0}^{3}&=\frac{1}{m_{\delta}^{2}}\underset{b}{\sum}
\left[g_{\delta b}\left(m_{\zeta b}^{\star}\right) \right.
\\ & \left.-\frac{g_{\delta b}}{m_b}
\left(m_{\zeta b}^{\star}\right)^{\frac{\zeta+1}{\zeta}}
\left(g_{\sigma b}\sigma_{0} + g_{\delta b}\delta_{0}^{3}\,I^{3b}+g_{\sigma^* b}\sigma_{0}^*\right)\right]I^{3b}\,\rho_{s b},\\
\phi_{0}&=\frac{1}{m_{\phi}^{2}}\underset{b}{\sum}g_{\phi b}\rho_{b}, \\
\sigma^*_{0}&=\frac{1}{m_{\sigma^*}^{2}}\underset{b}{\sum}
\left[g_{\sigma^* b}\left(m_{\zeta b}^{\star}\right) \right.
\\ & \left. -\frac{g_{\sigma^* b}}{m_b}
\left(m_{\zeta b}^{\star}\right)^{\frac{\zeta+1}{\zeta}}
\left(g_{\sigma b}\sigma_{0} + g_{\delta b}\delta_{0}^{3}\,I^{3b}+g_{\sigma^* b}\sigma_{0}^*\right)\right]\rho_{s b}, \\
\end{split}\end{equation}
where $\sigma_{0}$, $\omega_{0}$, $\varrho_{0}$, $\delta_{0}$, $\phi_{0}$ and $\sigma^*_{0}$ denote 
the classical expectation values of the meson fields.

As the density increases, it is more energetically favorable for the system to populate new degrees of freedom
in order to lower its Fermi energy. In particular, due to strong interaction processes,
hyperon species are predicted to start to become important at densities around $2\rho_0$, 
where $\rho_0$ is the nuclear saturation density.
Assuming that the matter is in $\beta$-equilibrium, one can easily verify that the many-body forces also
play a role in the particle threshold, as each particle species is populated beyond the following threshold:
\begin{equation}\label{population}
 q_{b_i} \mu^*_{n} - q_{e_i} \mu^*_{e} - g_{\omega b}\omega_0 - g_{\varrho b}I_{3 b} \varrho_{03} - g_{\phi b}\phi_0 \geq m_{\zeta b_i}^{*},
\end{equation}
where $q_{b_i}$ and $q_{e_i}$ represent the baryon and electric charges, respectively, 
and $\mu^*_n$ and $\mu^*_e$ are the neutron and electron effective chemical potentials, respectively. 

Assuming chemical equilibrium and isospin symmetry or charge neutrality, 
we calculate the equation of state for the model from the components of the stress-energy tensor. 
The pressure and energy density of the model have the standard expressions, with the introduction of the many-body contributions
to the effective mass of the baryons and the effective coupling constants of the scalar mesons.

Note that the inclusion of additional meson fields in the model has a direct effect on the behavior of matter at high densities.
The presence of the $\delta$ meson also affects asymmetric nuclear matter at 
saturation density. In the following sections, we analyze the results of the formalism proposed in this 
work for describing the properties of nuclear matter in both low and high densities regimes.
In order to study matter at high densities, we discuss the nucleon-hyperon interaction in what follows.


\subsection{Nucleon-hyperon interaction} \label{hyperon_coupling_section}

Since hyperons are not present in nuclear matter at saturation density and 
experimental data concerning their interaction are scarce in the literature, 
many authors in the past proposed models to describe the meson-hyperon coupling \cite{Moszkowski:1974gj, Glendenning:1991es, Pal:1999sq}. 
However, in the last decades various efforts were made in order to understand this sector of strong interactions
in more detail and, as a consequence, we have some experimental constraints regarding mainly the hyperon-nucleon interaction. 

In particular, the existence of bound $\Lambda$-hypernuclear states indicates an attractive potential $U_{\Lambda}^N=-30\, \mathrm{MeV}$
  at saturation \cite{Millener:1988hp}.
Concerning $\Sigma$-nucleon interactions, the absence of bound states in a survey for $\Sigma$ atoms \cite{Mares:1995bm,Bart:1999uh} 
and also scattering studies point towards a repulsive potential (for a review of the topic see ref. \cite{Friedman:2007zza} and references therein).
Investigations on quasi-free production of $\Xi$'s indicate an attractive potential of about $U_{\Xi}^N=-18\, \mathrm{MeV}$
  
\cite{Fukuda:1998bi,Khaustov:1999bz,SchaffnerBielich:2000wj} 
(for this topic one can check again ref. \cite{Friedman:2007zza}).
On the other hand, still little is known about the hyperon-hyperon interaction 
resulting from the limited knowledge of double-$\Lambda$ hypernuclei \cite{Gal:2003ze, Ahn:2013poa}.
The few experimental data point towards a weakly attractive $U_{\Lambda}^{\Lambda}$ potential \cite{Takahashi:2001nm,Gal:2011tb, Ahn:2013poa}. 
Nothing can be said about ${\Lambda}{\Xi}$ and ${\Xi}{\Xi}$ interactions.

In this work, we define the hyperonic coupling constants $g_{\omega Y}$, $g_{\varrho Y}$, $g_{\delta Y}$ and $g_{\phi Y}$ by using
the SU(6) spin-flavor symmetry \cite{Dover:1985ba,Schaffner:1993qj} for the vector mesons, described as follows:
\begin{equation}\begin{split}
\label{hys}
\frac{1}{3}g_{\omega N}= \frac{1}{2}g_{\omega \Lambda}=\frac{1}{2} g_{\omega \Sigma}=  g_{\omega \Xi}, \\
g_{\varrho N}=\frac{1}{2} g_{\varrho \Sigma}=g_{\varrho \Xi}, \quad g_{\varrho \Lambda} = 0, \\
-\frac{2\sqrt{2}}{3} g_{\omega N}= 2 g_{\phi \Lambda}= 2 g_{\phi \Sigma}= g_{\phi \Xi},
\end{split}\end{equation}
and assuming simple isospin scaling for the coupling of the $\delta$ meson:
\begin{equation}
g_{\delta N}=\frac{1}{2} g_{\delta \Sigma}=g_{\delta \Xi}, \quad g_{\delta \Lambda} = 0.
\end{equation}

Thus, the hyperonic couplings to the vector mesons are proportional to the number of strange quarks present inside each particle.
The rule for the isovector mesons is given by the proportion between the nucleon and the hyperon isospin. 
For example, since the $\Lambda$-hyperon is a singlet, it has zero isospin and, therefore,
does not couple to the $\varrho$ and $\delta$ mesons.

We obtain the hyperon-sigma coupling associated to the attractive interaction between hyperons and nucleons
by fitting the potential depths of the hyperons in nuclear matter \cite{Glendenning:1991es,Schaffner:1992sn}: 
\begin{equation}
\label{potentials}
U_Y^N= g_{\omega Y} \omega_0(\rho_0) - g_{\sigma Y} \sigma_0(\rho_0),
\end{equation}
following the values of \cite{SchaffnerBielich:2000wj}:
$U_{\Lambda}^N =-28\, \mathrm{MeV}$, $U_{\Sigma}^N=+ 30\, \mathrm{MeV}$ and $U_{\Xi}^N= -18\, \mathrm{MeV}$.

At first, we only consider the scalar-$\sigma\omega\varrho\delta\phi$ version of the model, for fixed values of 
the hyperon potentials. Then, in Section V, we come back to the role of the $\sigma^*$-meson 
introducing the $\sigma\omega\varrho\delta\phi\sigma^*$ version of the model in order to verify the effects
of coupling constants on the macroscopic properties of neutron stars.
In particular, we study the effects of the potential depths and of $g_{\sigma^* Y}$ on the maximum mass, radii and 
particle abundances of neutron stars. 


\section{Nuclear Matter Properties at Saturation} \label{saturation_section}

Our model should be in agreement with experimental data for the properties of saturated nuclear matter.
For this reason, we impose a saturation density of $\rho_0= 0.15 \,\mathrm{fm}^{-3}$
  and a binding energy per baryon of $B/A = -15.75\,\mathrm{MeV}$ \cite{Walecka1986} and
infer the nucleon-meson coupling constants by fitting the standard values of nuclear matter properties at saturation (details in the following).
The $\zeta$ parameter, associated with the many-body forces, is constrained to describe a realistic effective mass of the nucleon $m_N^*$ 
at saturation between $0.66-0.78 \,m_N$ \cite{Johnson:1987zza,Jaminon:1989wj}, a compressibility modulus $K_0$ between about $200-300 \,\mathrm{MeV}$ \cite{Blaizot:1980tw,Glendenning:1987gd,Sharma:1988zza}, 
a symmetry energy $a^0_{sym}$ between $25 - 35\,\mathrm{MeV}$ \cite{Tsang:2012se,Lattimer:2012xj} and a symmetry energy slope $L_0$ between $60-115 \,\mathrm{MeV}$ \cite{Chen:2005ti,Lattimer:2012xj,Lopes:2014wda}. 

At saturation, the isopin-symmetric nuclear matter has vanishing pressure, and is not populated by leptons nor hyperons (by definition). 
Also, due to the isospin symmetry, the mean values of the isovector mesons $\varrho$ and $\delta$ are zero.
To determine the constants $(g_{\sigma N}/m_{\sigma})^2$, $(g_{\omega N}/m_{\omega})^2$ and the effective mass of the nucleon $m^*_N$ at 
saturation, we solve the system of equations of zero pressure, experimental value of the
binding energy per nucleon and the $\sigma_0$ field equation of motion self-consistently. 

We calculate the compressibility modulus for symmetric nuclear matter, which is related to the curvature of the equation of state by:
\begin{equation}
\label{compressibility}
K_0= 9 \rho_0 \left[ \frac{d(\varepsilon/\rho)}{d\rho} \right]_{\rho=\rho_0},
\end{equation}
where $K_0$ corresponds to the value of the compressibility modulus at saturation density $\rho_0$ and $\varepsilon$ is the corresponding energy density of the system.
It is important to note that the $\zeta$-parameter relates the effective mass to the compressibility modulus \cite{Dexheimer:2007mt},
meaning that for each choice of $\zeta$, $K_0$ is calculated independently and is not used as an input to parameterize the model,
as is the case in most relativistic mean field models.

\begin{figure}
 \centering
 \vspace{1.0cm}
 \includegraphics[width=9.cm]{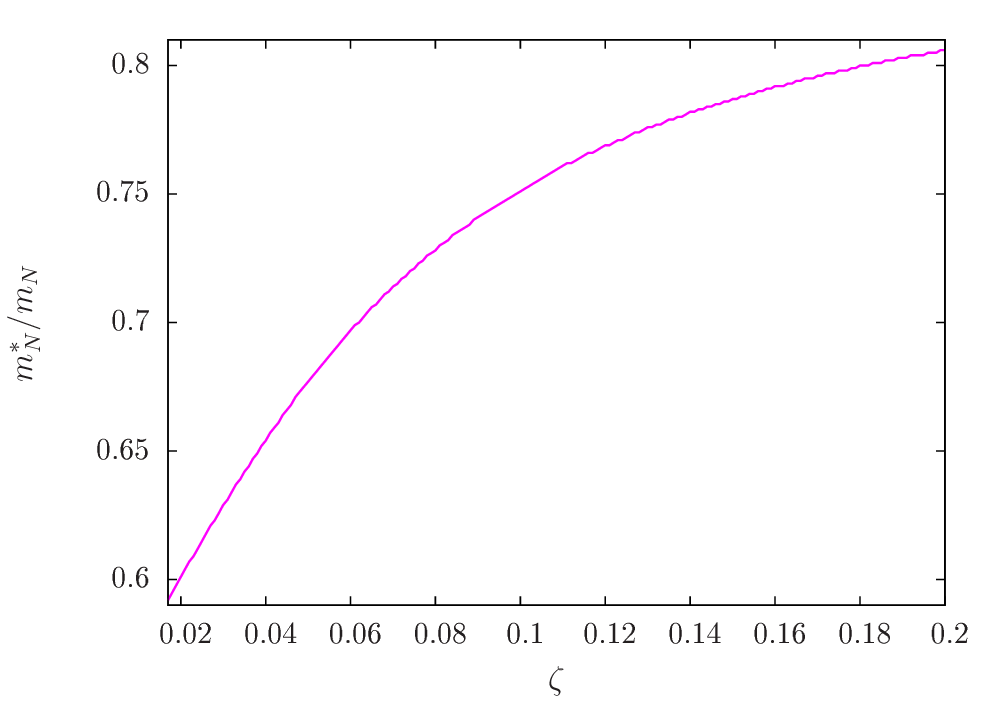}
 \caption{\label{meff_lambda} Effective mass of the nucleon at saturation densitiy as a function of the parameter $\zeta$.}
 \end{figure}

\begin{figure}
 \centering
 \vspace{1.0cm}
 \includegraphics[width=9.cm]{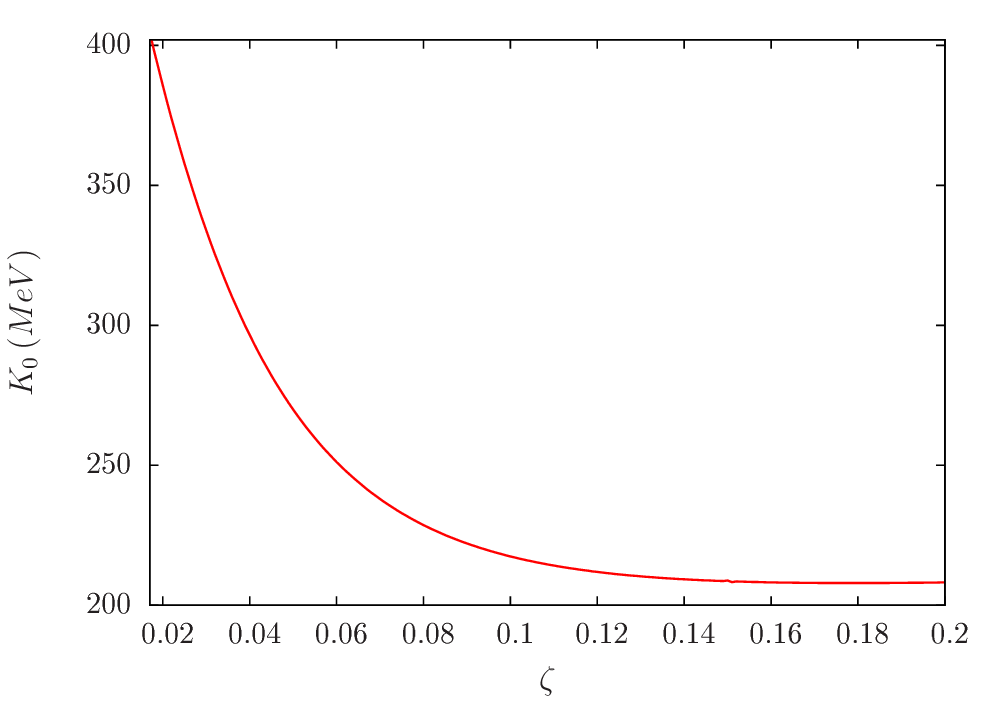}
 \caption{\label{K0lambda} Compressibility modulus at saturation density as a function of the parameter $\zeta$.}
 \end{figure}

\begin{figure}
 \centering
 \vspace{1.0cm}
 \includegraphics[width=9.cm]{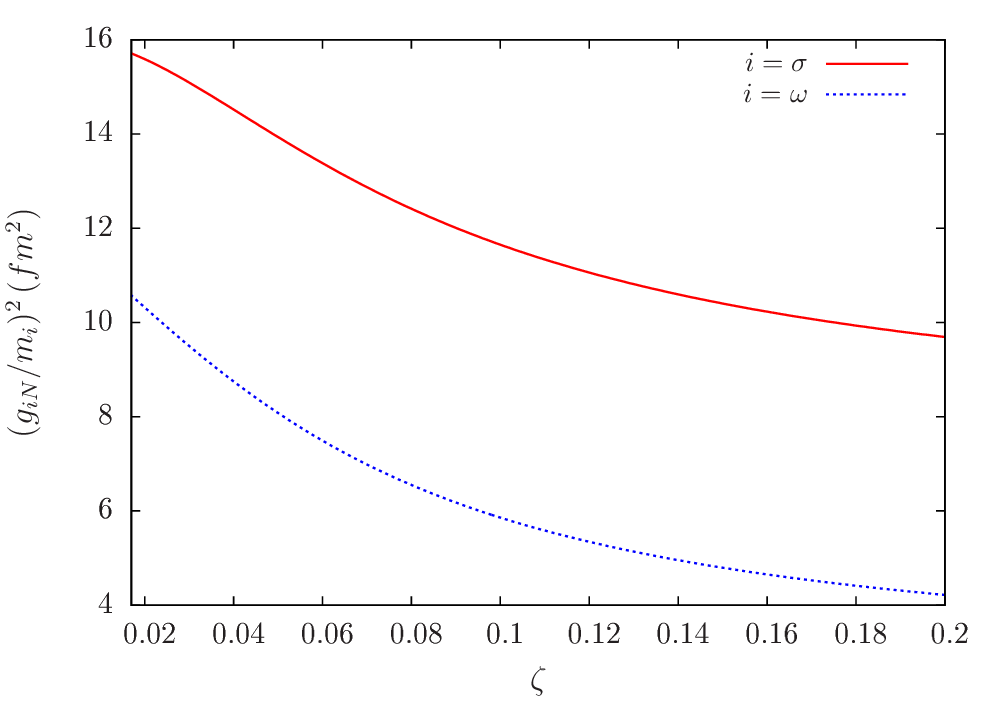}
 \caption{\label{coupling_sw_lambda} Coupling constants of the mesons $\sigma$ and $\omega$ as a function of the parameter $\zeta$.}
 \end{figure}

\begin{figure}
 \centering
 \vspace{1.0cm}
 \includegraphics[width=9.cm]{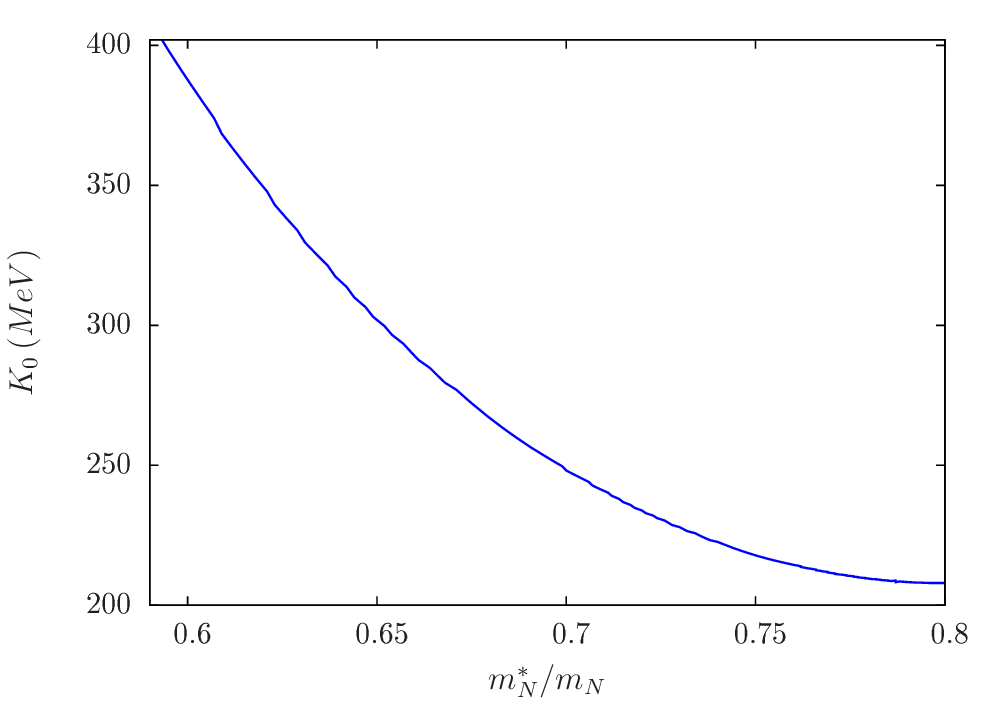}
 \caption{\label{meffK0} Direct relation between the effective mass of the nucleon and the compressibility modulus, 
 parametrized by $\zeta$ (at saturation density).}
 \end{figure}

The values of $(g_{\sigma N}/m_{\sigma})^2$, $(g_{\omega N}/m_{\omega})^2$, $m^*_N$ and $K_0$ for different 
choices of parameterizations (different $\zeta$) are shown in Table \ref{table:Table_coupling}.
Figures \ref{meff_lambda}, \ref{K0lambda} and \ref{coupling_sw_lambda} show 
that these quantities rapidly converge as a function of $\zeta$,
leaving a small range of values of $m^*_N$ and $K_0$ that fit the experimental values.
The direct relation between the effective mass and the compressibility modulus is shown in Figure \ref{meffK0}.
We remark that low values of the parameter $\zeta$ (lower effective mass and higher compressibility modulus) 
generate stronger coupling constants for the scalar $\sigma$-meson as well as for the vector $\omega$-meson.
The decrease of the attraction provided by the many-body forces (through the scalar mesons) together with the constant coupling of the 
vector mesons allows matter to be more repulsive for smaller values of the $\zeta$ parameter.
We also checked the dependence of the binding energy as a function of density on the many-body forces parameter,
from which we have found that larger values of the $\zeta$ parameter allow for more bound, i.e., more attractive matter.
This behavior of matter has a direct impact on the EoS for higher densities and, thus, on the
observational properties of neutron stars, as we discuss in Section V.

\begin{table}[t]
  \caption{\label{table:Table_coupling} Normalized effective mass of the nucleon, compressibility modulus
  and coupling constants for different parameterizations of the model (different $\zeta$'s).}
\begin{ruledtabular}
\begin{tabular}{ccccc}
$\zeta$ & $m^*_n/m_n$ &  $K_0$~(MeV) & $(g_{\sigma N}/m_{\sigma})^2$ & $(g_{\omega N}/m_{\omega})^2$  \\
  \hline
    &  &  &  &  \\
  0.040 & 0.66 & 297  & 14.51  & 8.74 \\ 
  0.045 & 0.67 & 282  & 14.22  & 8.40 \\ 
  0.049 & 0.68 & 272  & 13.99  & 8.14 \\    
  0.054 & 0.69 & 262  & 13.71  & 7.83 \\ 
  0.059 & 0.70 & 253  & 13.44  & 7.55 \\
  0.065 & 0.71 & 244  & 13.12  & 7.23 \\ 
  0.071 & 0.72 & 237  & 12.82  & 6.94 \\ 
  0.078 & 0.73 & 230  & 12.50  & 6.63 \\ 
  0.085 & 0.74 & 225  & 12.21  & 6.37 \\ 
  0.094 & 0.75 & 220  & 11.86  & 6.05 \\ 
  0.104 & 0.76 & 216  & 11.53  & 5.75 \\ 
  0.115 & 0.77 & 213  & 11.20  & 5.46 \\ 
  0.129 & 0.78 & 211  & 10.84  & 5.16 \\ 
  \end{tabular}
\end{ruledtabular}
\end{table}

 In order to determine the coupling constants of the nucleon with respect to the isovector mesons, 
 we must fit the properties of asymmetric nuclear matter.
 When only the $\varrho$ meson is considered, the respective coupling constant is fitted to the symmetry energy and
 the value of the slope of the symmetry energy at saturation is obtained directly from that.
 However, as pointed out by Lopes et. al. \cite{Lopes:2014wda}, the inclusion of the $\delta$ meson breaks this relation.
 In this case, it is necessary to consider the equation of state of asymmetric nuclear matter (in the absence of leptons)
 and solve the system of equations of the symmetry energy and its slope to find the corresponding values of $(g_{\varrho N}/m_{\varrho})^2$
 and $(g_{\delta N}/m_{\delta})^2$, according to:
 \begin{equation}
\label{symmetry}
a_{sym}^0= \frac{1}{2} \left[ \frac{d^2(\varepsilon/\rho)}{dt^2} \right]_{t=0},
\quad
L_0= 3 \rho_0 \left[ \frac{d a_{sym}}{d \rho} \right]_{\rho=\rho_0},
\end{equation}
where the asymmetry between protons and neutrons is quantified by $t=(\rho_p-\rho_n)/\rho_b$.

It is important to note that, in this formalism, the many-body contributions correlate the effective mass of the nucleon and 
the scalar field equations in the presence of the delta meson (see equation \ref{mstar_zen}). 
In other words, to determine the equation of state of asymmetric matter at saturation, it is 
necessary to first solve self-consistently a system of equations for the expressions of $m^*_N(m^*_N,,\sigma_0, \delta_0)$, 
$\sigma_0(m^*_N,\sigma_0, \delta_0)$  and $\delta_0(m^*_N,\sigma_0, \delta_0)$
for a non-vanishing isospin system.

Although the value of the slope of the symmetry energy at saturation density has been widely discussed 
in the literature, its values still lie in a large accepted range.
Since some studies point towards values around $60 \,\mathrm{MeV}$ \cite{Chen:2005ti,Tsang:2012se} and others towards higher values even as high as $113 \,\mathrm{MeV}$ \cite{Steiner:2011ft,Lattimer:2012xj}, 
we perform a large scan of values with our model in order to find the isovector-mesons coupling constants suitable
to describe massive hyperon stars.
 
\begin{figure}[ht]
\centering
  \includegraphics[width=9cm]{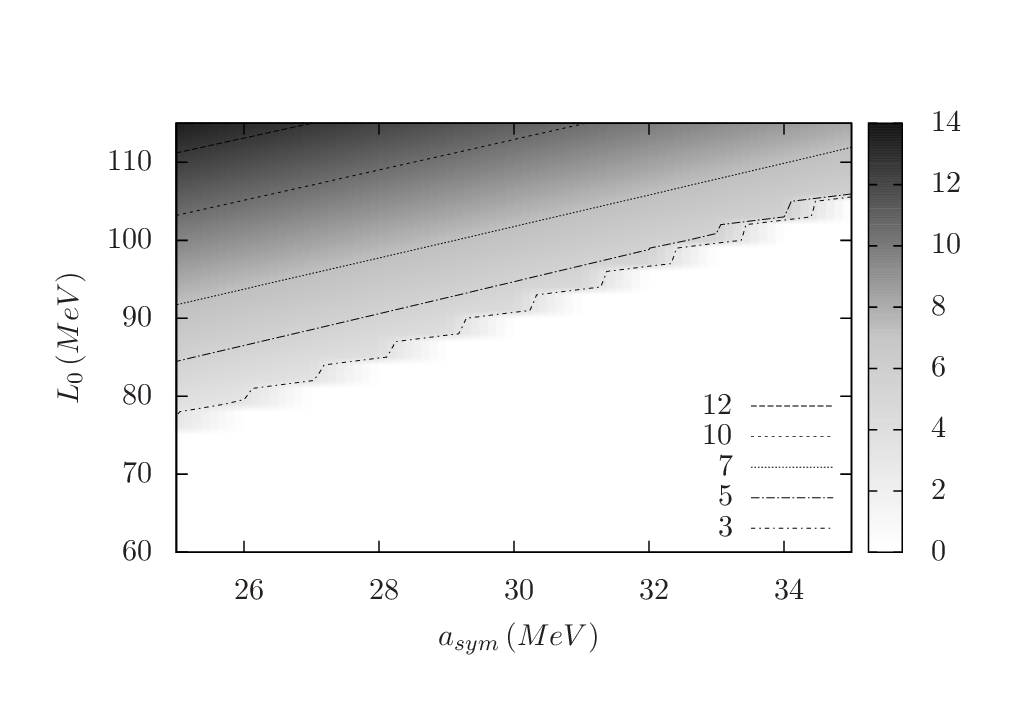}
      \centering
  \includegraphics[width=9cm]{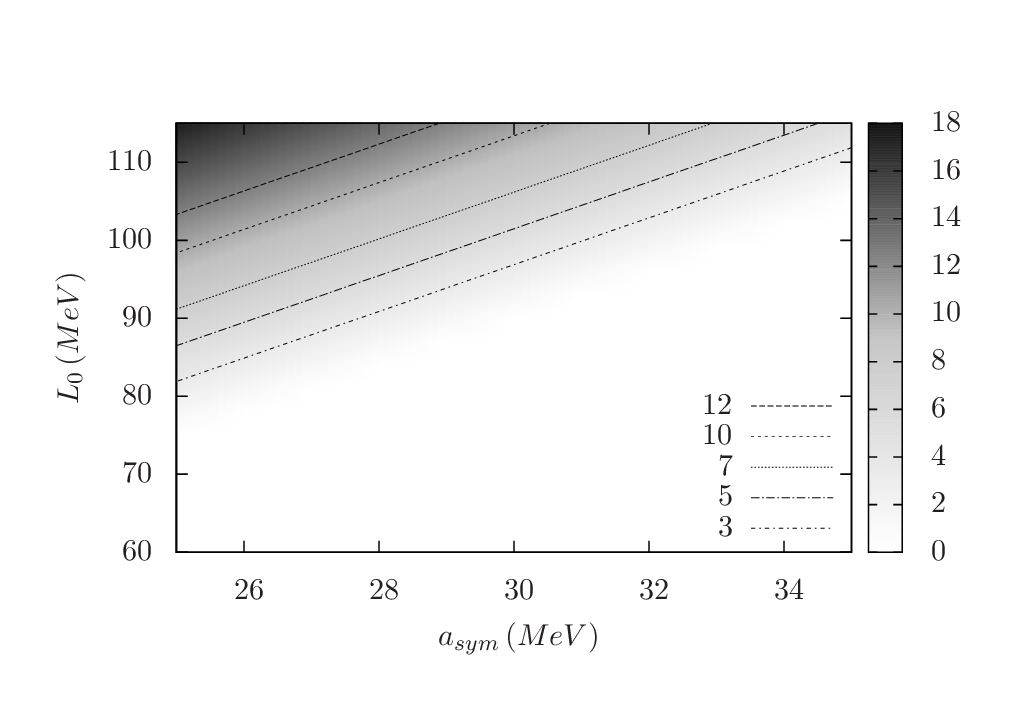}
      \caption{Possible values of $a_{sym}^0$ and $L_0$ for $(g_{\varrho N}/m_{\varrho})^2$ (top panel) and 
      $(g_{\delta N}/m_{\delta})^2$ (bottom panel) for $\zeta=0.040$. The intensity of each coupling is plotted
      in a color sequence and also indicated by the different types of lines. }\label{map_l0040}
\end{figure}

\begin{figure}[ht]
\centering
  \includegraphics[width=9cm]{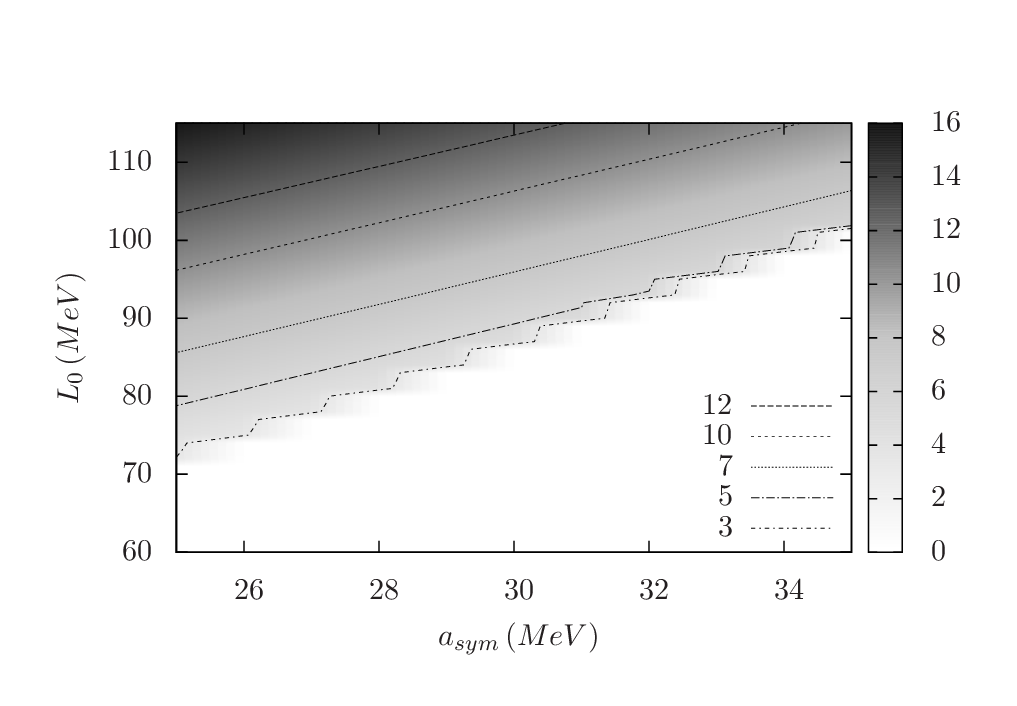}
    \centering
  \includegraphics[width=9cm]{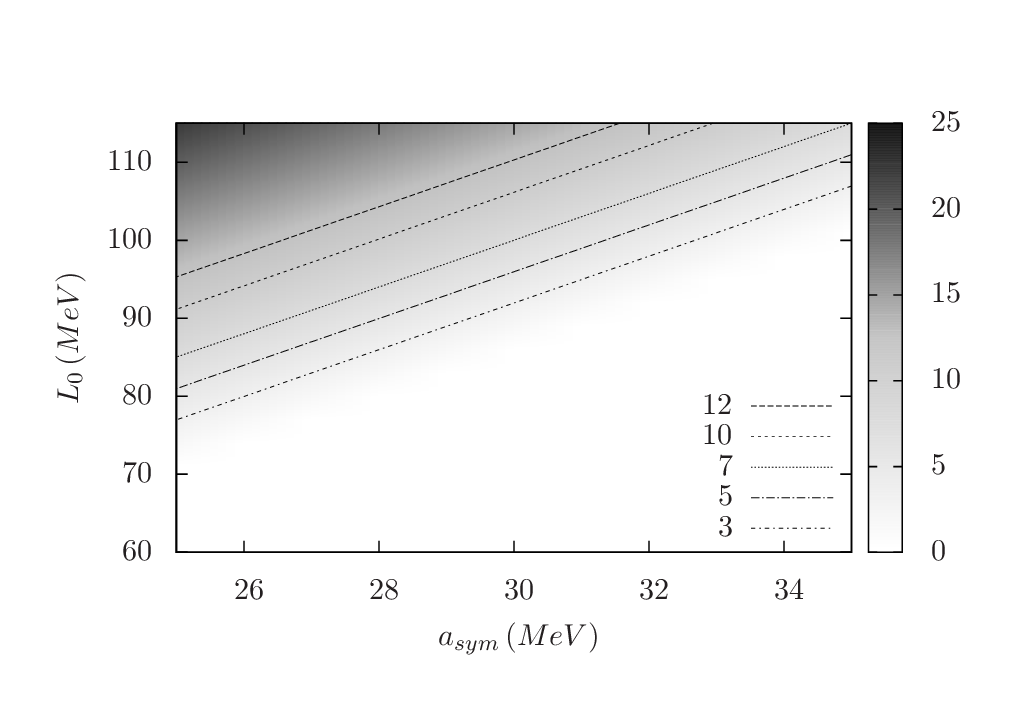}
    \caption{Same as Figure \ref{map_l0040} for $\zeta=0.071$. }\label{map_l0071}
\end{figure}

\begin{figure}[ht]
\centering
  \includegraphics[width=9cm]{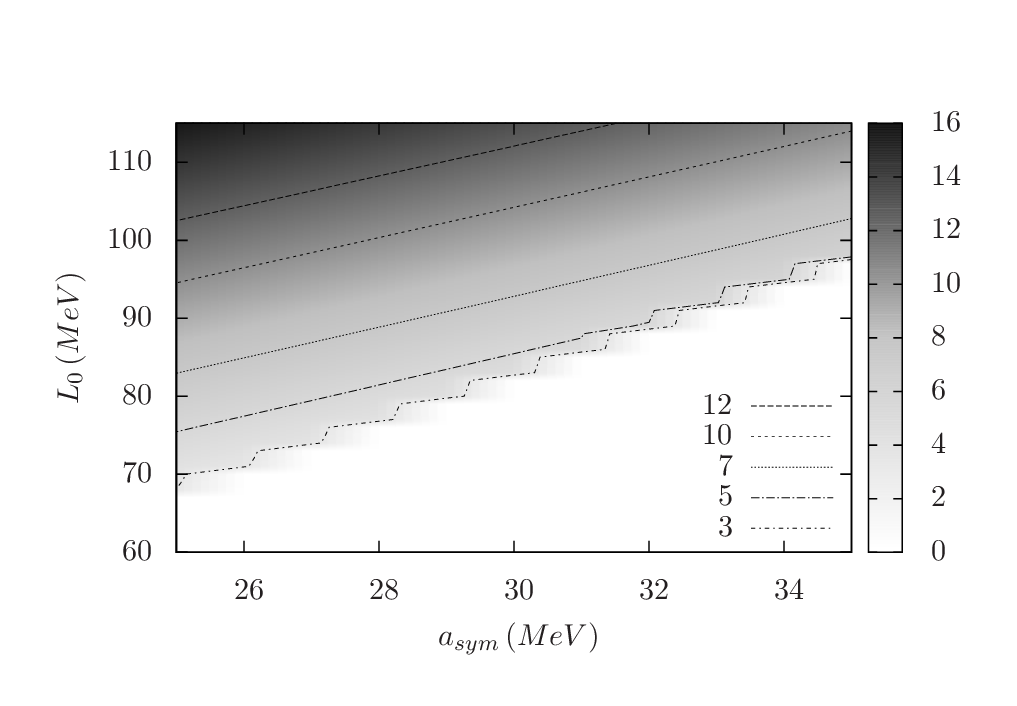}
    \centering
  \includegraphics[width=9cm]{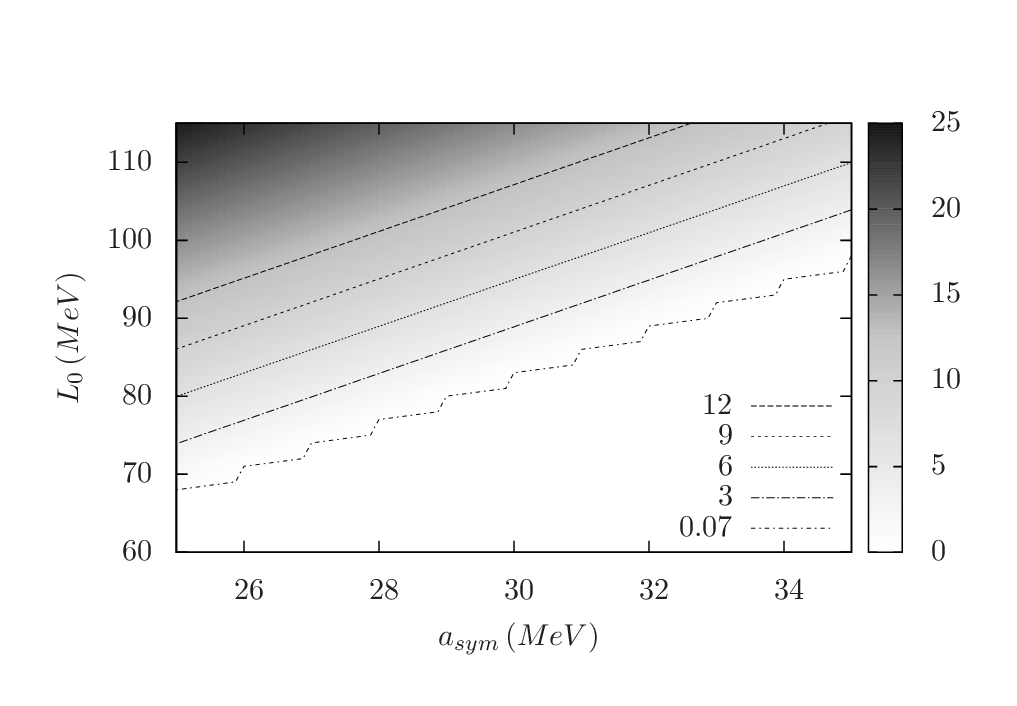}
    \caption{Same as Fig \ref{map_l0040} for $\zeta=0.129$. }\label{map_l0129}
\end{figure}

In summary, the methodology used to describe nuclear properties in this formalism is the following:
by the analysis of the properties of symmetric matter at saturation, we choose the non-linearity parameter $\zeta$
that will determine the $g_{\sigma N}$ and $g_{\omega N}$ coupling constants and give the values of the 
effective mass of the nucleon and the compressibility modulus by a one-to-one relation.
However, since the choice of the parameter $\zeta$ is not enough to determine
$g_{\varrho N}$ and $g_{\delta N}$ uniquely, we must analyze the new parameter space given by the isovector meson coupling constants,
the symmetry energy and its slope as shown in Figures \ref{map_l0040}, \ref{map_l0071} and \ref{map_l0129} for different choices of $\zeta$.

Differently from the case of symmetric matter, where for a given value of the effective nucleon mass there is only one possible value for the compressibility modulus
(Figure \ref{meffK0}), Figures \ref{map_l0040}, \ref{map_l0071} and \ref{map_l0129} show that for a given value of the symmetry energy $a_{sym}^0$,
one can have different values of the slope $L_0$. The choices of $a_{sym}^0$ and $L_0$ result in different values of the coupling constants of the mesons
$\varrho$ and $\delta$, and are also dependent on the parameter $\zeta$.

The parameter space for $\zeta=0.040$ is shown in Figure \ref{map_l0040}. The panels show the coupling constants in a color scale as a 
function of $a_{sym}^0$ and $L_0$. The color scale on the top panel corresponds to the values 
of $(g_{\varrho N}/m_{\varrho})^2$. The same is shown for $(g_{\delta N}/m_{\delta})^2$ on the bottom panel.
The white regions in the figures correspond to the cases where no solution for the system of equations exists.
Obviously, the values of the coupling constants must be selected simultaneously in both panels, in order to guarantee that their
values are associated with the same solutions for $a_{sym}^0$ and $L_0$.

Note that, by the comparison between different choices of the parameter $\zeta$, there is a substantial change in the possible 
range of solutions for the same values of the coupling constants. 
For example, comparing the bottom panels in Figures \ref{map_l0040} and \ref{map_l0129}, one verifies that the solution for 
$(g_{\delta N}/m_{\delta})^2 = 12 \mathrm{fm}^{2}$
  is found in the interval of $a_{sym}^0\simeq 25-27\, \mathrm{MeV}$
  and $L_0\simeq 111-115\, \mathrm{MeV}$
  
for $\zeta=0.040$, while for $\zeta=0.129$ the interval is increased to $a_{sym}^0\simeq 25-31.5\, \mathrm{MeV}$
  and $L_0\simeq 102-115\, \mathrm{MeV}$
 .
This analysis was carried out only in the range of $25 \leq a_{sym}^0\leq 35\, \mathrm{MeV}$
  and $60 \leq L_0 \leq 115 \, \mathrm{MeV}$
  in this work. 

The maps for the symmetry energy and its slope show that the minimum values found for the asymmetry energy slope are those that correspond 
to the higher values of the parameter $\zeta$. The lowest value for the slope can be seen in Figure \ref{map_l0129}, 
where for $a_{sym}^0=25 \,\mathrm{MeV}$, the solution of $L_0=68$ is associated with $(g_{\varrho N}/m_{\varrho})^2=2.96\,\mathrm{fm}^{2}$
  (top panel) and
$(g_{\delta N}/m_{\delta})^2=0.07\,\mathrm{fm}^{2}$ (bottom panel).

\begin{figure}[ht]
\centering
   \includegraphics[width=9.cm]{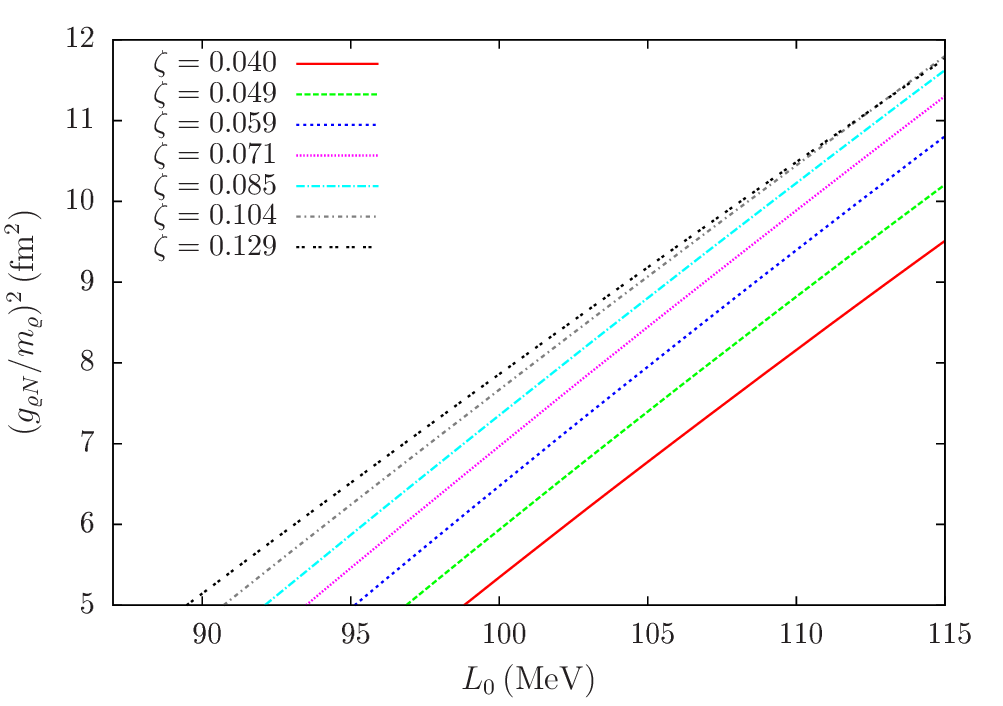}
    \centering
   \includegraphics[width=9.cm]{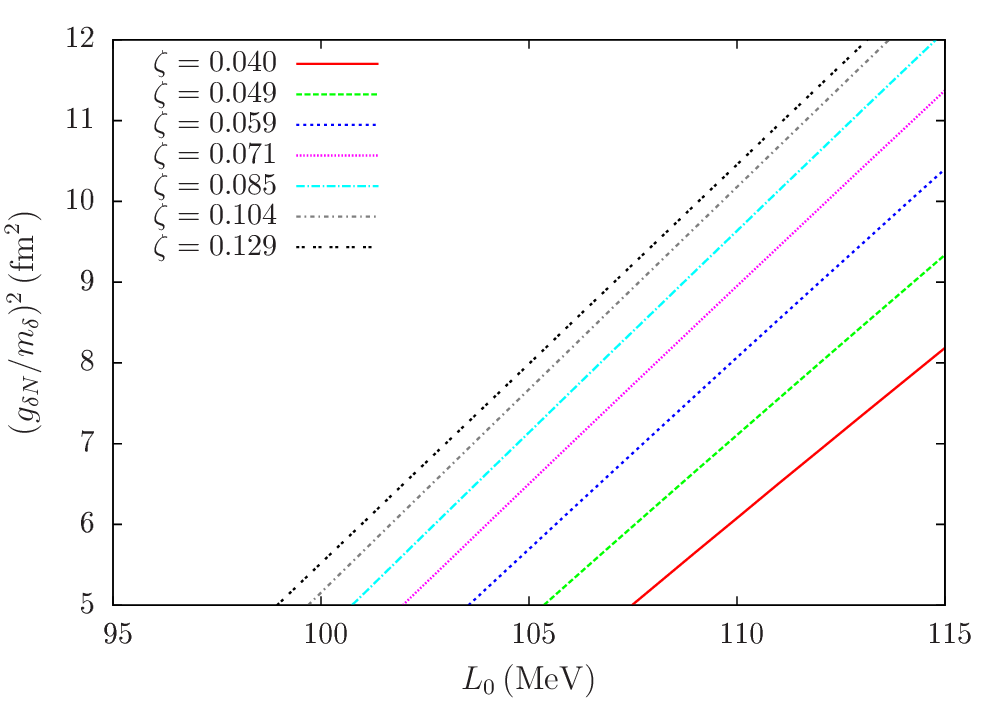}
 \caption{\label{gdL0_lambda} $(g_{\varrho N}/m_{\varrho})^2$ (top panel) and $(g_{\delta N}/m_{\delta})^2$ (bottom panel)
 dependence on the slope of the symmetry energy $L_0$ for 
  different value of the $\zeta$ parameter. 
  The plots correspond to the symmetry energy $a_{sym}^0=32\, \mathrm{MeV}$
 .}
 \end{figure}

Figure \ref{gdL0_lambda} allows us to better visualize the relation between the coupling constants intensity and the slope $L_0$
for different choices of $\zeta$.
The top and bottom panels show simultaneous solutions for $(g_{\varrho N}/m_{\varrho})^2$ and $(g_{\delta N}/m_{\delta})^2$, 
from which we verify a linear relation with respect to the slope of the symmetry energy $L_0$. 
The higher values of the parameter $\zeta$ are those which provide lower $L_0$ values, but higher 
coupling constants $(g_{\varrho N}/m_{\varrho})^2$ and $(g_{\delta N}/m_{\delta})^2$.
In particular, this result is important in the light of recent works pointing towards low values of $L_0$ \cite{Hebeler:2013nza},
meaning that the $\delta$ meson contribution should not be too large.

A recent work by Dutra et al. that carries out a review of 263 parameterizations of different RMF models reported that the value of the volume
part of the isospin incompressibility, denoted $K_{\tau v}$, can also be used as a constraint for the description of nuclear matter \cite{Dutra:2014qga}.
This quantity reads:
\begin{equation}
\label{K_tauv}
K^0_{\tau v}=  \left(K^0_{sym} - 6 L_0 - \frac{Q_0}{K_0}L_0   \right),
\end{equation}
where $K_0$ and $L_0$ are the compressibility modulus and the slope of the symmetry energy at saturation.
The quantities $Q_0$ and $K^0_{sym}$ are the skewness coefficient of the equation of state and the curvature of the symmetry
energy, respectively: 
\begin{equation}\begin{split}
\label{derivatives}
Q_0=  27\rho_0^3 \left( \frac{d^3(\varepsilon/\rho)}{d\rho^3}  \right)_{\rho=\rho_0, t=0}, \\
K^0_{sym}=  9\rho_0^2 \left( \frac{d^2(a_{sym})}{d\rho^2}  \right)_{\rho=\rho_0}.
\end{split}\end{equation}

The values of $K^0_{\tau v}$ and $Q_0$ lay in a wide range of uncertainties, $K^0_{\tau v}=-550 \pm150 \, \mathrm{MeV}$
 
and $Q_0=-700 \pm 500\, \mathrm{MeV}$, which come from the overlap of the analysis of isospin diffusion \cite{Chen:2004si}, 
neutron skin \cite{Centelles:2008vu} and measurement of Sn isotopes \cite{Li:2007bp}.
We have also calculated the symmetry energy skewness at saturation
$Q^0_{sym}=  27\rho_0^3 \left( \frac{d^3 a_{sym}}{d\rho^3}  \right)_{\rho=\rho_0}$,
and the values of all these quantities for different parameterizations of the model are shown in Table \ref{Tab_properties}.

We find that the parameterizations $\zeta=0.040,\,0.049,\,0.059,\,0.071$ present values of $K^0_{\tau v}$ in agreement with the literature. 
The results for the skewneess coefficient $Q_0$ for the parameterizations $\zeta=0.040$ and $\zeta=0.049$, are smaller than the threshold value.
However, we must emphasize that the quantities analyzed in Table \ref{Tab_properties} have a wide range of uncertainties and are based on the overlap of 
experimental data that carry large uncertainties themselves. 
Also, the calculation of these quantities depends directly on the values of the symmetry energy and its slope and 
the results presented in Table \ref{Tab_properties} are for fixed values of the $a_{sym}^0$ and $L_0$.
A wider study regarding these quantities is out of the scope of this present work.

\begin{table}[t]
\caption{Volume part of the isospin incompressibility $K_{\tau v}$, curvature of the symmetry energy $K^0_{sym}$,  
skewness coefficient of the symmetry energy $Q^0_{sym}$, and of the equation of state $Q_0$ at saturation 
for different parameterizations of the model. 
The values of the symmetry energy and its slope are fixed to $a_{sym}^0=32\, \mathrm{MeV}$
  and $L_0=97\, \mathrm{MeV}$
 .} 
\begin{center} \label{Tab_properties}
\begin{tabular}{|c|c|c|c|c|} 
\hline 
$\zeta$ & $K^0_{\tau v}(\mathrm{MeV})$ & $K^0_{sym}(\mathrm{MeV})$ & $Q^0_{sym}(\mathrm{MeV})$ & $Q_0(\mathrm{MeV})$ \\
& & & & \\ 
\hline \hline
$0.040 $ &  $-558$ & $30.85$ & $184.5$ & $21.3$ \tabularnewline
$0.049 $ &  $-484$ & $31.23$ & $197.8$ & $-188.2$ \tabularnewline
$0.059 $ &  $-412$ & $30.72$ & $192.3$ & $-364.6$ \tabularnewline
$0.071 $ &  $-346$ & $29.30$ & $176.1$ & $-505.9$ \tabularnewline
$0.085 $ &  $-291$ & $27.36$ & $150.4$ & $-612.1$ \tabularnewline
$0.104 $ &  $-243$ & $25.01$ & $114.8$ & $-698.8$ \tabularnewline
$0.129 $ &  $-214$ & $22.57$ & $78.3$ & $-751.1$ \tabularnewline
\hline\hline
\end{tabular}
\end{center}
\end{table}



\section{Properties of Asymmetric nuclear Matter  at high densities} \label{prop_high_section}
Before turning our attention to neutron star matter, we discuss the behavior of asymmetric matter properties at high densities.
The symmetry energy and its slope can be used to extrapolate the description of nuclear matter 
to isospin asymmetric nuclear matter at higher densities.
Also, as was already pointed out by Lopes et. al. \cite{Lopes:2014wda} and other authors, the behavior of the properties
of asymmetric nuclear matter in the high density regime has a significant impact on the observable properties  of neutron stars. 

Allowing matter to be populated by nucleons and hyperons, we assume conserved isospin and baryon number and start by calculating 
the compressibility for symmetric nuclear matter as a function of the density, using the 
general expression:
\begin{equation}
\label{K_rho_eq}
K (\rho)=  9 \left(\frac{dP}{d\rho} \right)_{t=0}.
\end{equation}

\begin{figure}
 \centering
 \vspace{1.0cm}
 \includegraphics[width=9.cm]{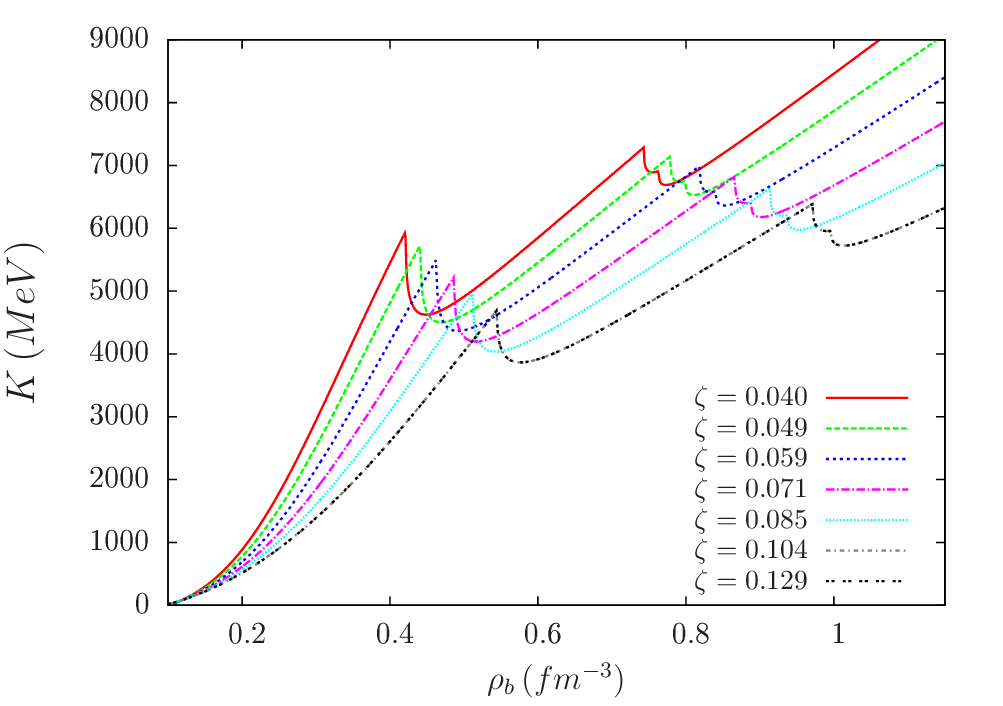}
 \caption{\label{k_rho} Compressibility modulus of asymmetric matter (containing hyperons) as a function of baryon density for different choices of $\zeta$.}
 \end{figure}

The behavior of the compressibility as a function of density is shown in Figure \ref{k_rho} for different
choices of the parameter $\zeta$ and for fixed values of the hyperon potentials 
($U_{\Lambda}^{N}=-28MeV,\,U_{\Sigma}^{N}=30MeV,\,U_{\Xi}^{N}=-18\, \mathrm{MeV}$
 ), 
symmetry energy $a^0_{sym}=32\, \mathrm{MeV}$
  and slope $L_{0}=97\, \mathrm{MeV}$
  at saturation.
The results shown in Figure \ref{k_rho} point to the fact that the lower values of $\zeta$ generate 
higher values of the compressibility for high densities, 
analagously to the behavior at saturation density.

The impact of the appearance of new degrees of freedom can be verified at densities around
$0.5\, \mathrm{fm}^{-3}$ and $0.7\,\mathrm{fm}^{-3}$, which correspond to the densities where the $\Lambda$ and $\Xi$
hyperons appear.
As the density increases and new degrees of freedom are populated, the compressibility modulus
decreases due to the softening of the EoS (when a hyperon appears). 
The density continues to increase and the Fermi momentum of the new particle species increases,
making the EoS stiff again, until another particle appears. 

\begin{figure}
 \centering
 \vspace{1.0cm}
 \includegraphics[width=9.cm]{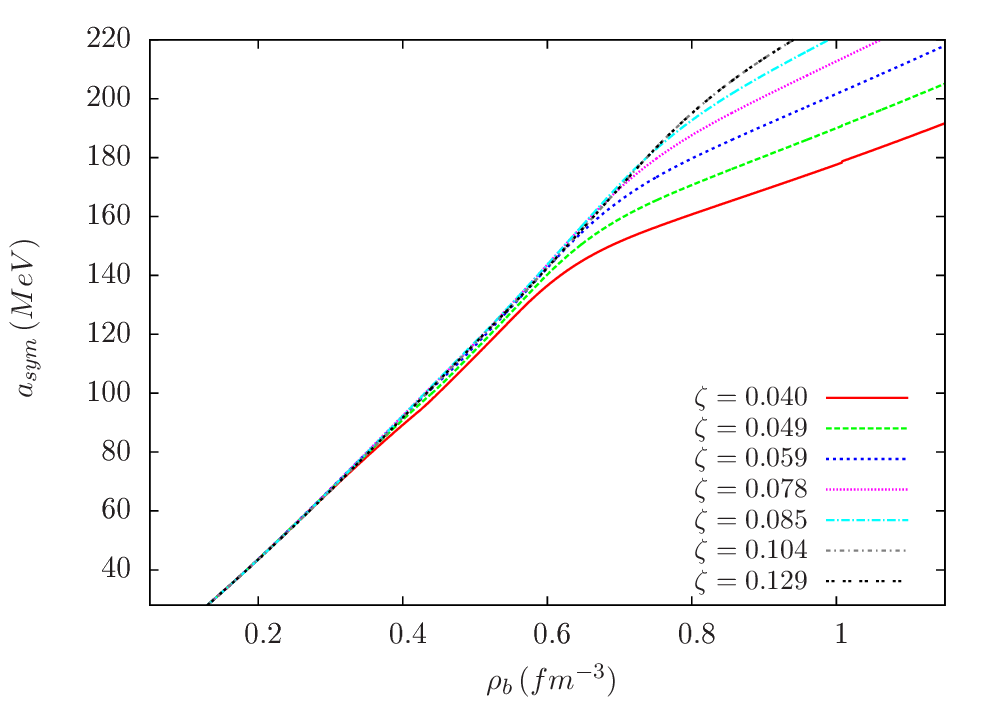}
 \caption{\label{asym_rho} Symmetry energy of hyperonic matter as a function of the baryon density for different choices of $\zeta$.}
 \end{figure}

\begin{figure}
 \centering
 \vspace{1.0cm}
 \includegraphics[width=9.cm]{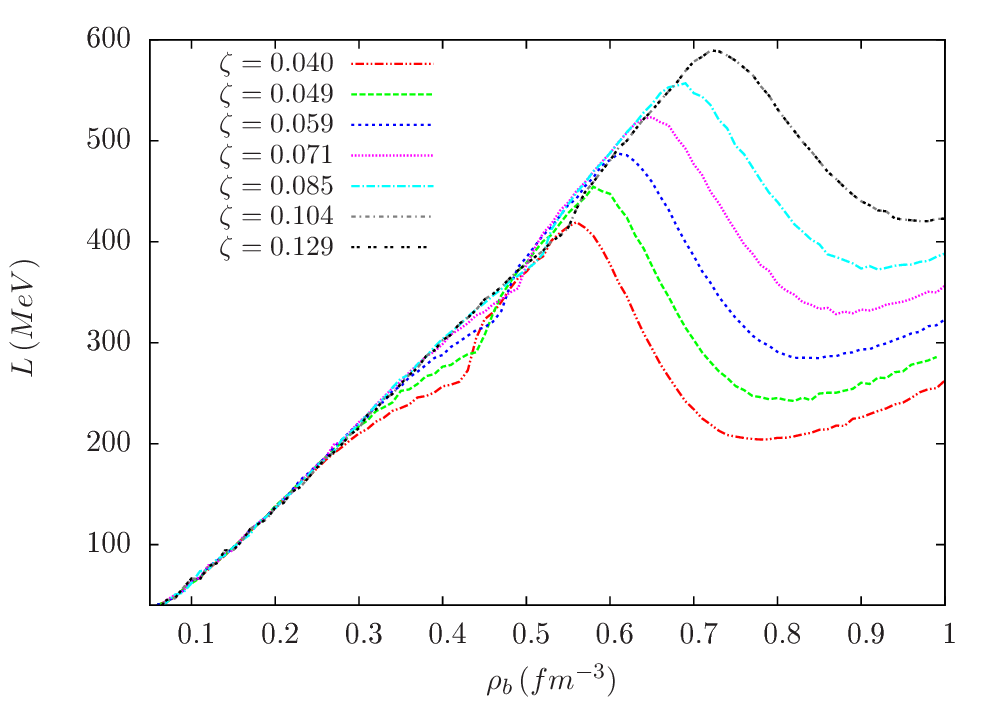}
 \caption{\label{L_rho} Slope of the symmetry energy of hyperonic matter as a function of the baryon density for different choices of $\zeta$.}
 \end{figure}

 We calculate the symmetry energy and its slope by extending the expressions in equation \ref{symmetry} to higher densities.
 The respective results are shown in Figures \ref{asym_rho} and \ref{L_rho} and correspond, again, to different choices 
 of the parameter $\zeta$ at fixed values of the hyperon potentials, symmetry energy $a^0_{sym}=32\, \mathrm{MeV}$
  and slope $L_{0}=97\, \mathrm{MeV}$
  at saturation.
 Figure \ref{asym_rho} shows that the symmetry energy is not affected by the parameter $\zeta$ at low densities,
 as expected from our previous fitting, and starts to
 present an interesting behavior only for densities of about $~3 \rho_0$, which correspond to the point where hyperons (appear for symmetric matter).
 
 For fixed values of $a^0_{sym}$ and its slope $L_0$ at saturation, higher values of the $\zeta$ parameter introduce larger differences between symmetric and 
 asymmetric matter EoS's. 
 This behavior is due to the fact that the coupling constants of the isovector mesons are also dependent on the values of the $\zeta$ parameter, generating a competition
 between the attraction and repulsion from the isoscalar and isovector mesons.
 In particular, higher values of $\zeta$ generate stronger couplings of the $\varrho$ meson, which contribute only to the asymmetric EoS
 and are of importance at high densities.
 The strongest repulsive contribution provided by the higher values of the $\zeta$ parameter (higher $\varrho$ couplings) allows for a stiffer EoS that differs more from the 
 symmetric EoS (in the absence of the isovector mesons contributions) and, since the symmetry energy relates precisely this difference, the parameterization with higher 
 values of $\zeta$ yields higher values of the symmetry energy as a function of density.
 
 Since the symmetry energy and the slope of the symmetry energy are correlated quantities, 
 it is natural that they present a similar behavior:
 the higher (lower) the $\zeta$ parameter, the higher (lower) are the values of the symmetry energy and its slope as a function of the density.
 The peaks in Figure \ref{L_rho} are again due to the appearance of the hyperons.
 
 It is worth mentioning that different values of the symmetry energy $a_{sym}^0$ and its slope $L_0$ at saturation also have 
 an impact on their behavior at higher densities. In particular, smaller values of $a_{sym}^0$ and higher values of 
 $L_0$ allow for a higher increase in the slope. We will come back to this topic in next section,
 when the impact of these properties on the radii of neutron stars is investigated. 
 
\begin{figure}
\centering
 \includegraphics[width=9.cm]{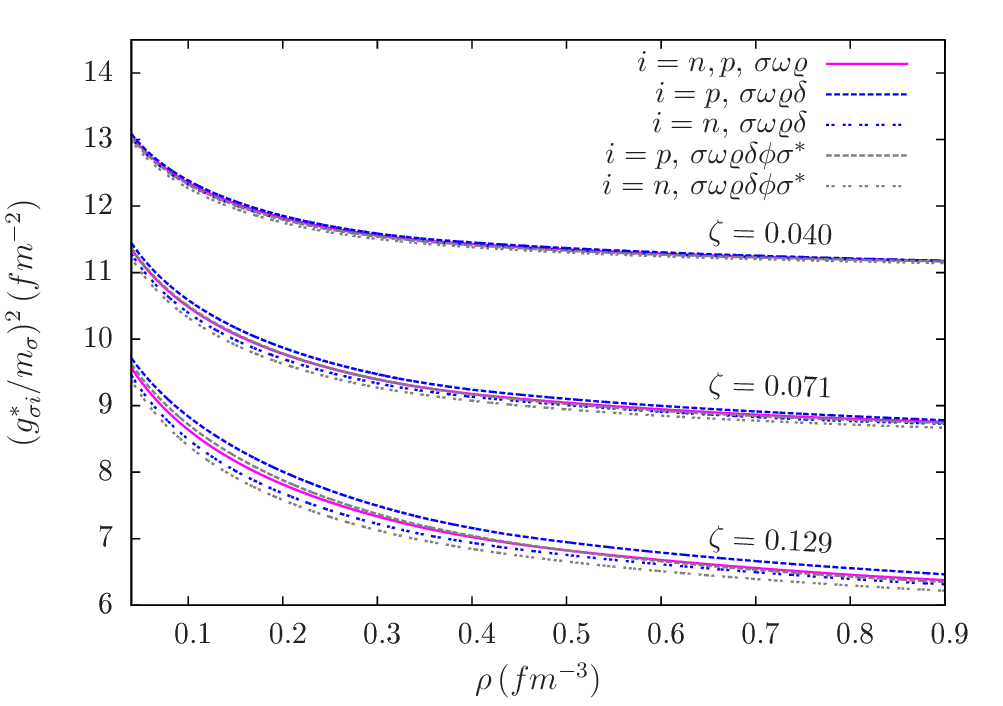}
     \centering
 \includegraphics[width=9.cm]{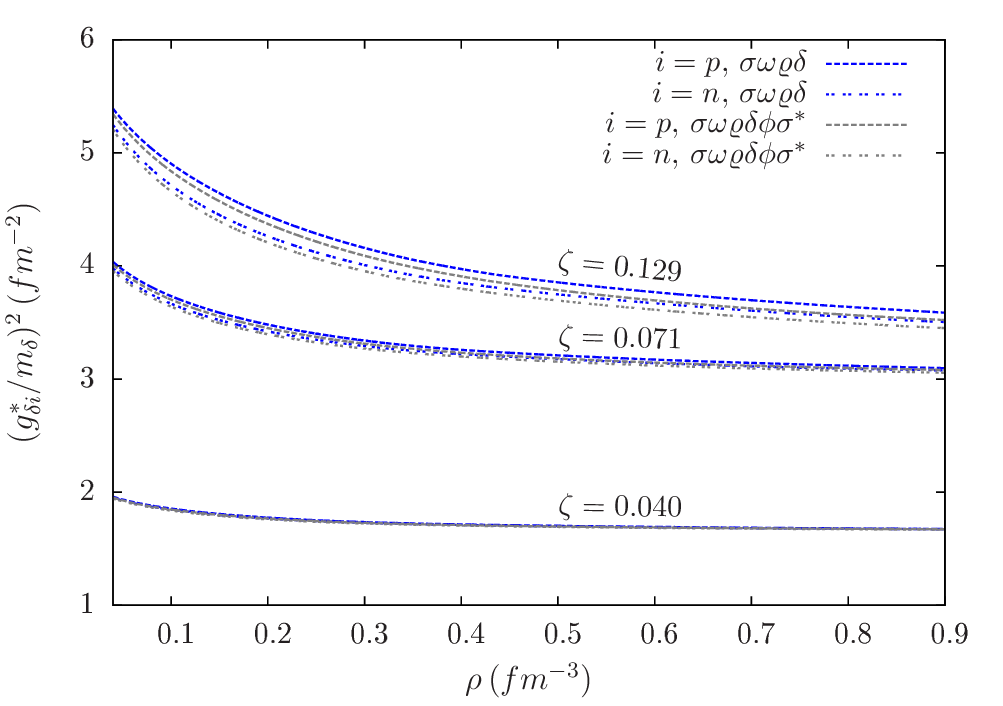}
 \includegraphics[width=9.cm]{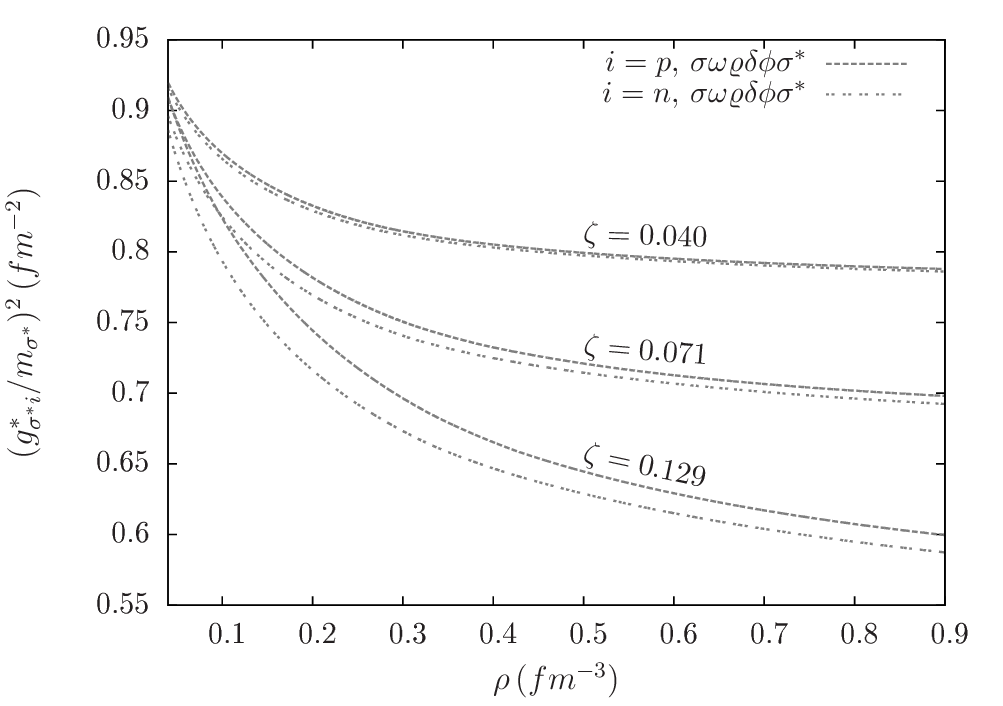}
     \caption{Effective coupling constants of the scalar mesons as a function of baryon density.
     The coupling of the $\sigma$ (top panel), $\delta$ (middle panel) and $\sigma^*$ (bottom panel) are plotted
     for protons and neutrons for different values of $\zeta$. The different versions of the model are indicated by colors: 
     $\sigma\omega\varrho$ (magenta), $\sigma\omega\varrho\delta$ (blue) and $\sigma\omega\varrho\delta\phi\sigma^*$ (grey).} \label{geff_rho}
\end{figure}

\begin{figure}
 \centering
 \vspace{1.0cm}
 \includegraphics[width=9.cm]{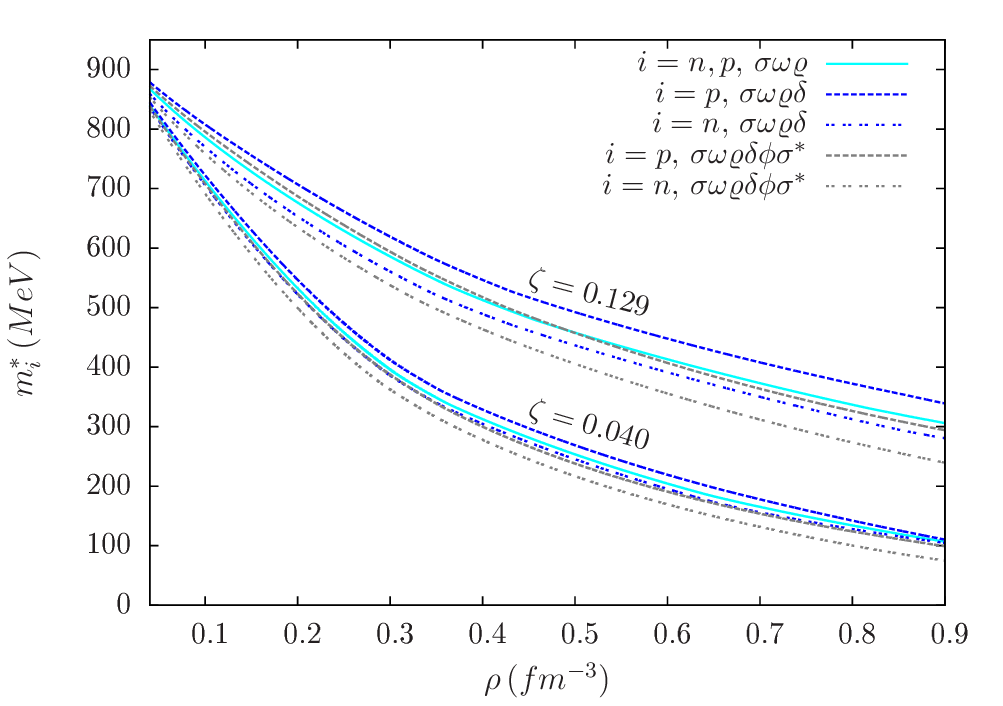}
 \caption{\label{meff_rho} Effective mass of protons and neutrons as a function of baryon density for $\zeta=0.040$ and $\zeta=0.129$.
 Each color correspond to a different version of the model.}
 \end{figure}

Finally, we mention that, although the meson fields are density dependent in every RMF model,
in the formalism adopted in this work the coupling between baryons and scalar mesons depends 
on the nonlinear contributions of the meson fields themselves. 
This feature of the model implies that the behavior of the scalar mesons effective coupling constants
($g_{\sigma b}^*$, $g_{\delta b}^*$ and $g_{\sigma^* b}^*$) as a function of the density is affected 
by the many-body contributions, generating a direct impact on the baryon effective masses and,
consequently, on the global behavior of the matter. 
This behavior is quantified in Figures \ref{geff_rho} and \ref{meff_rho}, where we
show the effective coupling constants of the scalar mesons and the effective mass
of the nucleons as a function of baryon density.
We show results for different versions of the model (with different meson content)
as well as different choices of the $\zeta$ parameter.

In the top panel of Figure \ref{geff_rho}, we show that the introduction of the $\delta$ meson
breaks the isospin degeneracy in the coupling constants of protons and neutrons.  
The difference is more pronounced for higher values of $\zeta$, which correspond to
stronger many-body contributions. 
The delta meson increases the positive isospin particles coupling with respect to the 
scalar mesons and decreases the coupling for negative isospin particles, 
meaning that the first generates more attraction. 
On the other hand, from Equation \ref{tcm}, one can see that the isospin of the particles affects
all scalar mesons non-trivially due to the nonlinear contributions.
Hence, the competition between the amount of particles with positive and negative isospin
plays an important role for the global attractive or repulsive response of matter.

The introduction of the $\sigma^*$ meson has impact on all of the scalar field equations as well.
The top and middle panels of Figure \ref{geff_rho} show that this meson contributes to a faster decrease of the effective coupling constants,
due to its extra contribution to the many-body forces that change the couplings.
Also, from the bottom panel of Figure \ref{geff_rho} one can see that
the proton and neutron coupling constants departure from the degenerate case is more notable
when all scalar fields, together with higher values of the many-body forces contribution, are assumed.

The effective mass of the nucleon as a function of the density in Figure \ref{meff_rho} 
presents a similar response to the introduction of the $\delta$ and $\sigma^*$ mesons.
The $\delta$ meson splits the masses of protons and neutrons and the $\sigma^*$ meson makes 
their decrease faster due to the extra contribution in the many-body forces.
Both effects are extremely relevant to the chemical equilibrium and, consequently, to the
global behavior of the matter. 
In the next section, we will come back to the effects of the $\sigma^*$ meson for neutron star matter with hyperons.

\section{Astrophysical Application } \label{astro_application_section}

In this section, we apply the formalism developed so far to describe hyperon matter inside neutron stars.
We compare our predictions with the recently observed massive neutron stars PSR J038+0432 and PSR J1614-2230 \cite{Demorest2010,Antoniadis2013}.
We have already shown that our model is in agreement with nuclear matter properties at saturation density, 
and now a comparison with astrophysical properties characterizes a second test of the underlying microscopic model.
In addition, aiming to verify the impact of future hypernuclear data in the context of neutron stars,
we investigate the effects of different values of the hyperon potentials and the coupling constant of 
the $\sigma^*$-meson on macroscopic properties of neutron stars.

\begin{figure}
 \centering
 \vspace{1.0cm}
 \includegraphics[width=9.cm]{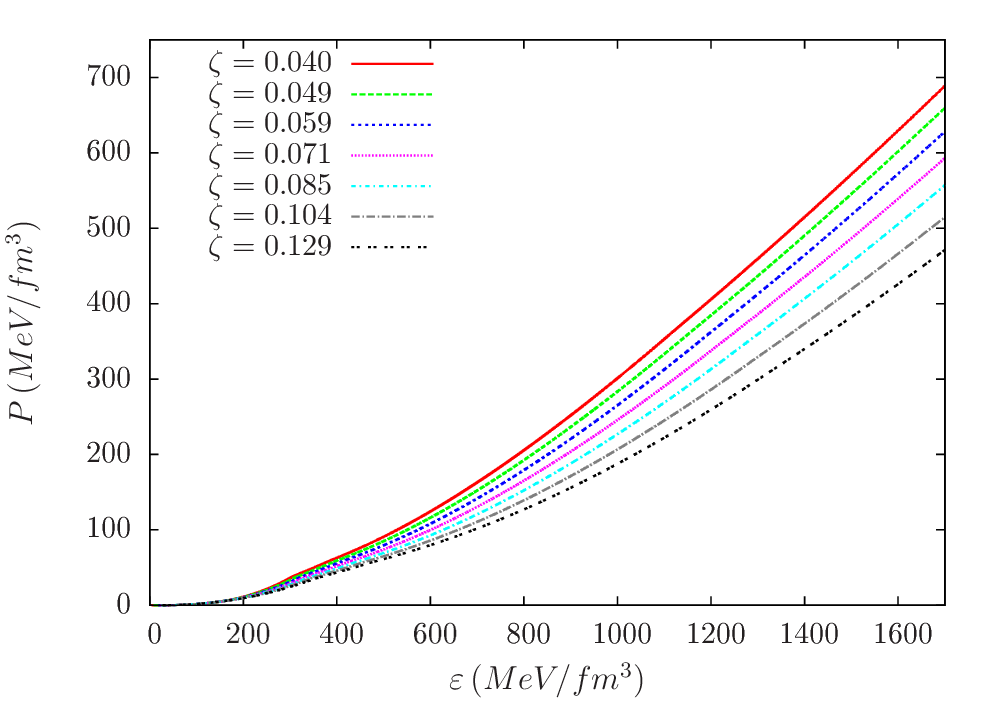}
 \caption{\label{eos_lambda} Equation of state for star matter shown for different parameter sets.}
 \end{figure}

\begin{figure}
 \centering
 \vspace{1.0cm}
 \includegraphics[width=9.cm]{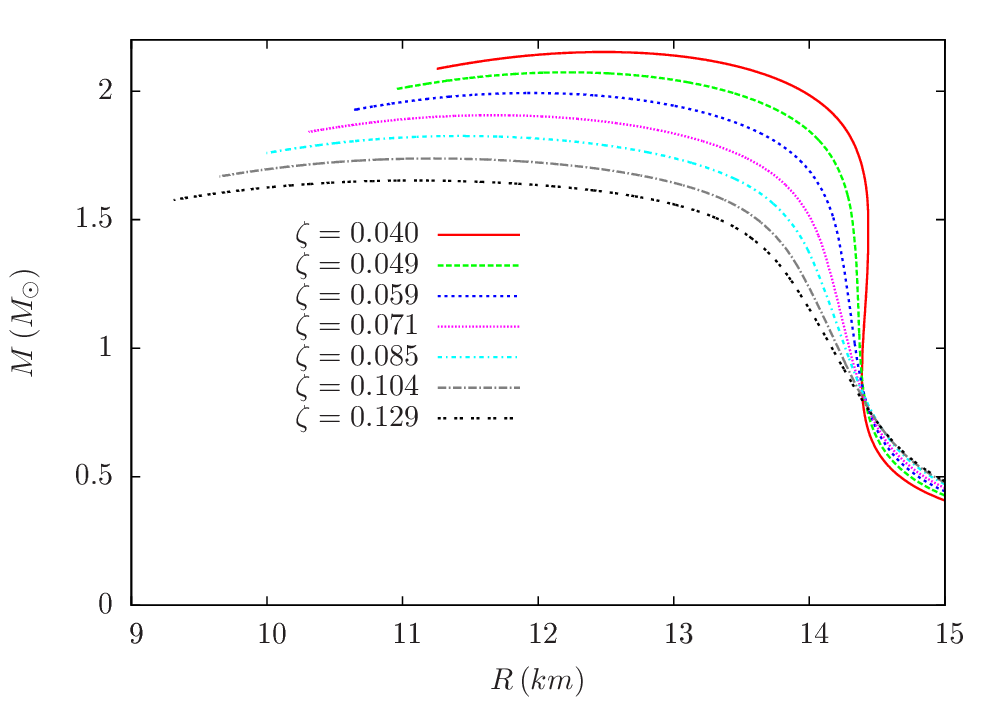}
 \caption{\label{tov_lambda} Mass-Radius relation for star matter shown for different sets of parameters.}
 \end{figure}

Assuming that matter is in $\beta$-equilibrium and is locally charge neutral, we use the EoS of the model as an input to 
solve the {\it Tolman-Oppenheimer-Volkoff} (TOV) equations \cite{Tolman1939,Oppenheimer1939}  and
obtain the macroscopic structure of stars for different parameterizations.
The EoS and the Mass-Radius diagram for each parameterization studied, including hyperons, can be found 
in Figures \ref{eos_lambda} and \ref{tov_lambda}, respectively.


 \subsection{The role of $\zeta$ on the properties of hyperon star} \label{zetastar_section}

The dependence of the EoS on the parameters of the model is shown in Figure \ref{eos_lambda}. 
In particular, the lowest value of $\zeta$ generates the stiffer 
EoS and, consequently, allows for a higher stellar mass. 
The argument here is the same as the one discussed in Section III: 
since a stiffer EoS is related to
higher values of the internal pressure of the system and,
accordingly, to higher values of 
the compressibility modulus of nuclear matter, this in turn requires stronger
contributions from repulsive components of 
the nuclear force. 
In our general approach, however,
many-body forces lower the strenghs both of attractive and
repulsive interaction terms due to {\it shielding effects}, which
result in higher (lower) values of the compressibility modulus
of nuclear matter in the case of higher (lower)
relative reduction of the attractive contributions. 
Our present results are consistent with the
analysis above, since we have obtained a stiffer EoS
for lower values of the 
$\zeta$ parameter, which corresponds to stronger repulsion. 

In summary, lower values of the $\zeta$ parameter
generate more repulsive nuclear matter which is able to support more gravity and, 
consequently, create a macroscopic object with higher mass. 
In particular, the highest stellar mass generated by the extended version of model including hyperons is $2.15 M_{\odot}$ (for $\zeta = 0.040$), 
in agreement with pulsars PSR J1614–2230 ($M = 1.97 \pm 0.04 M_\odot$) \cite{Demorest2010},
and PSR J0348+0432 ($M = 2.01 \pm 0.04 M_\odot$) \cite{Antoniadis2013}, as is shown in Figure \ref{tov_lambda}.

For completeness, in Table \ref{table:Table_modelmass} we present results of different versions of the model. 
The first and second columns represent different meson content versions of the model and values of the $\zeta$ parameter, respectively.
We also show maximum mass results for nucleonic and hyperon stars, in the third and fourth columns, respectively.
Also, since the $\sigma\omega\varrho$ version of the model has a one-to-one relation between 
$a_{sym}^0$ and $L_0$, we show values of $L_0$ in the fifth column.

All nucleonic stars in the range of parameters of $\zeta=0.040-0.129$ are in agreement with observational data.
In particular, the most massive nucleonic star described by the model is the one for $\zeta=0.040$, with $2.57M_{\odot}$.
We also verify that the inclusion of the $\delta$ meson does not affect significantly the maximum mass of the stars,
introducing a difference of only $0.01M_{\odot}$ in the model. The weak contribution of the $\delta$ meson in the 
results comes from the low values of $g_{\delta N}$ chosen in order to reproduce low values of the slope $L_0$.

The $\delta$ meson introduces a repulsion between protons and neutrons.
As this repulsion favors the neutron population, the Fermi momenta of neutrons harden the
EoS, generating more massive nucleonic stars.
In the case of hyperon stars, the $\delta$ meson shifts the appearance of hyperons to lower densities,
preventing a strong stiffening of the EoS and even softening it in some cases.
Therefore, in general, for the nucleonic stars, the $\delta$ meson has the effect of increasing the maximum star mass,
while for hyperon stars, the meson decreases its maximum value \cite{Menezes:2004vr}.

Assuming that hyperons are present in compact stars, ensuring that the model is in accordance with
the observational data puts additional constraints on nuclear properties at saturation.
The comparison between the parameterizations in Tables \ref{table:Table_coupling} and \ref{table:Table_modelmass}
shows that the version of the model presented in this work constrains the effective mass of the nucleon and the compressibility modulus 
on ranges of $0.66-0.70\,m_N$ and $253-297\, \mathrm{MeV}$
  (associated to the range of $\zeta=0.040-0.059$), respectively. 
Also, from the parameterization $\zeta=0.059$, the lowest values of the slope in agreement with nuclear and
observational data is $L_0=94\,\mathrm{MeV}$ (for $a_{sym}^0=32\,\mathrm{MeV}$).
Such a conclusion can be reached because a change in the slope $L_0$ from $97\,\mathrm{MeV}$ to $115\,\mathrm{MeV}$ only contributes to
a tiny decrease of less that $0.0035M_{\odot}$ in the maximum mass of stars.
We also find a decrease of $0.01M_{\odot}$ in the maximum star mass when the symmetry energy $a_{sym}$ is changed from $30\,\mathrm{MeV}$ to $33\,\mathrm{MeV}$.

One can also clearly see from Figure \ref{tov_lambda} that the radius of the star is affected by the $\zeta$ parameter.
We will come back to this topic in subsection \ref{radius_section}, where we investigate all quantities that affect the 
radius of the canonical star.

\begin{table}[t]
  \caption{\label{table:Table_modelmass} Maximum masses of nucleonic and hyperon stars for different versions of the model, $L_0$ and $\zeta$-parameter. The hyperon potentials are fixed to $U_{\Lambda}^N=-28 \,\mathrm{MeV}$, $U_{\Sigma}^N=+30\,\mathrm{MeV}$
   $U_{\Xi}^N=-18 \,\mathrm{MeV}$ and the symmetry $a_{sym}^0=32\,\mathrm{MeV} $.}
\begin{ruledtabular}
\begin{tabular}{ccccc}
Model & $\zeta$ & $M^{nucleonic}_{max}\,(M_{\odot})$ &  $M^{hyperon}_{max}\,(M_{\odot})$ & $L_0\,(\mathrm{MeV})$  \\
  \hline
    & &  & & \\
$\sigma\omega\varrho$ &   0.040 & 2.57 & 1.90 & 96.16  \\ 
                      &   0.049 & 2.49 & 1.83  & 94.68  \\    
                      &   0.059 & 2.41 & 1.76  & 93.22  \\
                      &   0.071 & 2.32 & 1.69  & 92.05  \\ 
                      &   0.085 & 2.24 & 1.62  & 90.92  \\ 
                      &   0.104 & 2.15 & 1.55  & 89.82  \\ 
                      &   0.129 & 2.06 & 1.49  & 88.86  \\ 
  
  \hline
      & &  & &  \\
$\sigma\omega\varrho\delta$ &   0.040 & 2.57 & 1.90  & 97.0  \\ 
                            &   0.049 & 2.50 & 1.83  & 97.0  \\ 
                            &   0.059 & 2.42 & 1.76  & 97.0  \\ 
                            &   0.071 & 2.33 & 1.68  & 97.0  \\  
                            &   0.085 & 2.25 & 1.61  & 97.0  \\  
                            &   0.104 & 2.16 & 1.55  & 97.0  \\ 
                            &   0.129 & 2.07 & 1.49  & 97.0  \\  
  
  \hline
      & &  &  &  \\
$\sigma\omega\varrho\delta\phi$ &   0.040 & 2.57 & 2.15  & 97.0  \\ 
                                &   0.049 & 2.50 & 2.07  & 97.0  \\     
                                &   0.059 & 2.42 & 1.99  & 97.0  \\                        
                                &   0.071 & 2.33 & 1.91  & 97.0  \\                         
                                &   0.085 & 2.25 & 1.83  & 97.0  \\  
                                &   0.104 & 2.16 & 1.74  & 97.0  \\                         
                                &   0.129 & 2.07 & 1.65  & 97.0  \\  

  \end{tabular}
\end{ruledtabular}
\end{table}
  
\begin{figure}
\centering
\includegraphics[width=9.cm]{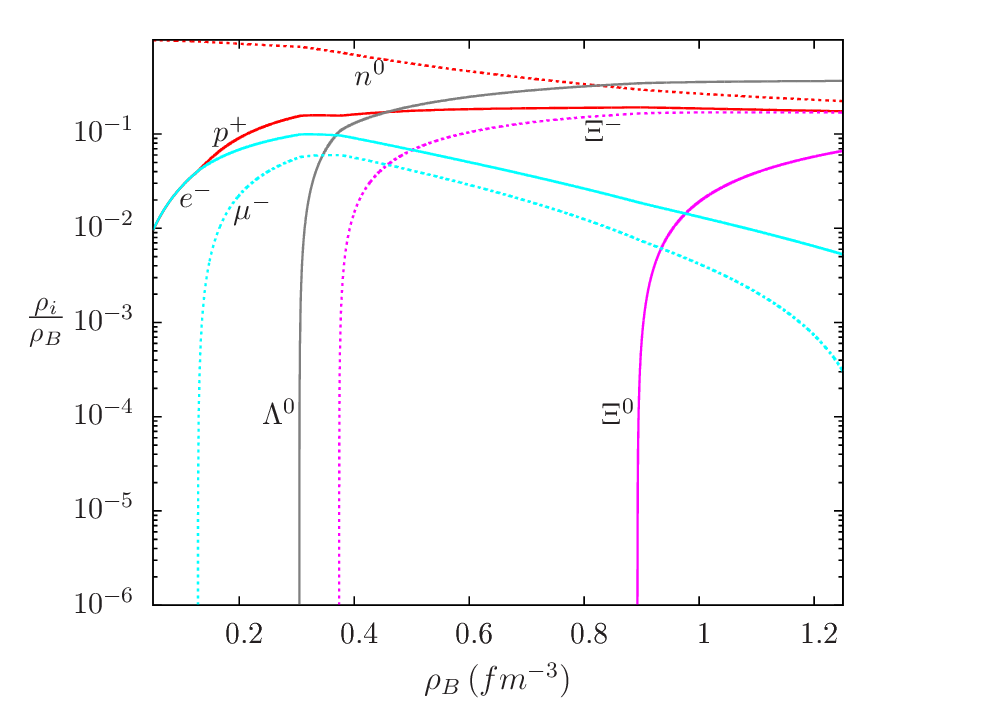}
    \centering
  \includegraphics[width=9.cm]{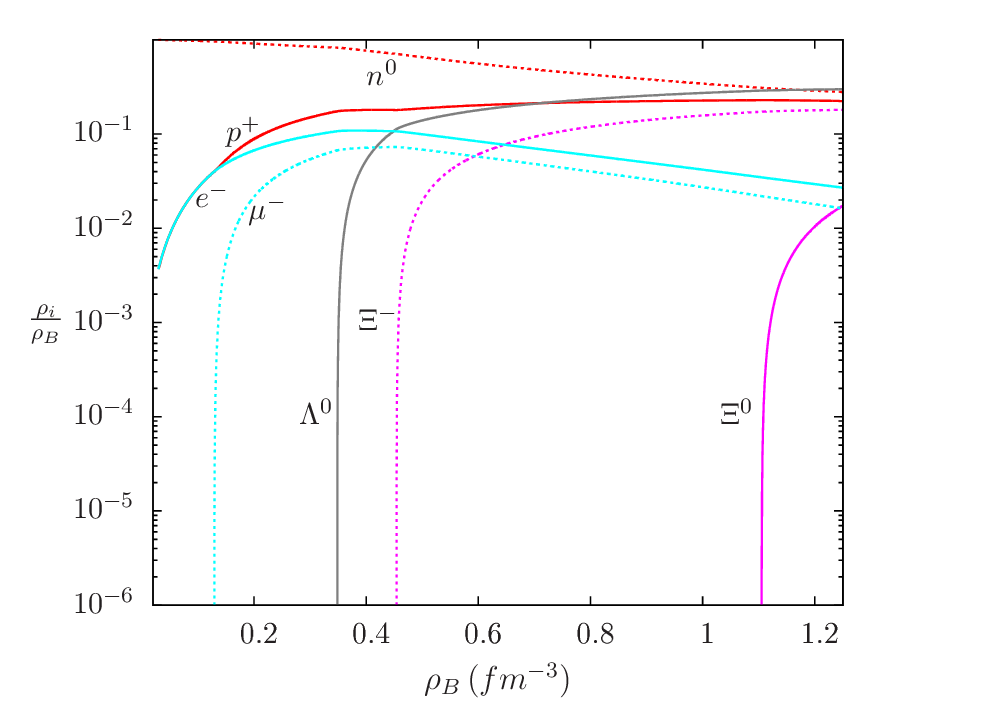}
    \caption{Particle population dependence on the parameter $\zeta$. The top panel shows the
    population for $\zeta=0.040$ and the bottom for $\zeta=0.129$. The x-axis represents the 
    baryon density and the y-axis the fraction of each particle specie 
    normalizeed by the total baryon density.}\label{pop_lambda}
 \end{figure}

In the last section, it was discussed that the many-body contributions present in the scalar meson 
couplings have a direct influence on the behavior of particles in hyperonic matter,
since they affect the chemical equilibrium equations. 
Also, since the $\zeta$ parameter dictates the strength of the nonlinear contributions, 
we analyze its effects on the particle population in Figure \ref{pop_lambda}.
The figure shows the fraction of particles as a function of baryon density for
two choices of parameters and for fixed values of the hyperon potentials, 
symmetry energy $a^0_{sym}=32\, \mathrm{MeV}$, and slope $L_{0}=97\, \mathrm{MeV}$
  at saturation.   

The results in Figure \ref{pop_lambda} show that, as the value of $\zeta$ increases, the densitiy corresponding to 
the appearance of each hyperon is shifted to higher values.
For the choices $\zeta=0.040$ (top panel) and $\zeta=0.129$ (bottom panel), the thresholds of the $\Lambda^0$, $\Xi^-$ and $\Xi^0$ hyperons
are shifted from $~0.30$ to $0.36\,\mathrm{fm}^{-3}$, $~0.38$ to $0.45\,\mathrm{fm}^{-3}$
  and $~0.90$ to $1.1\,\mathrm{fm}^{-3}$, respectively.
One can also verify that the threshold for the $\mu^-$ lepton is not altered by the parameter,
since it controls only the baryon interaction.

\begin{figure}
 \centering
 \vspace{1.0cm}
 \includegraphics[width=9.cm]{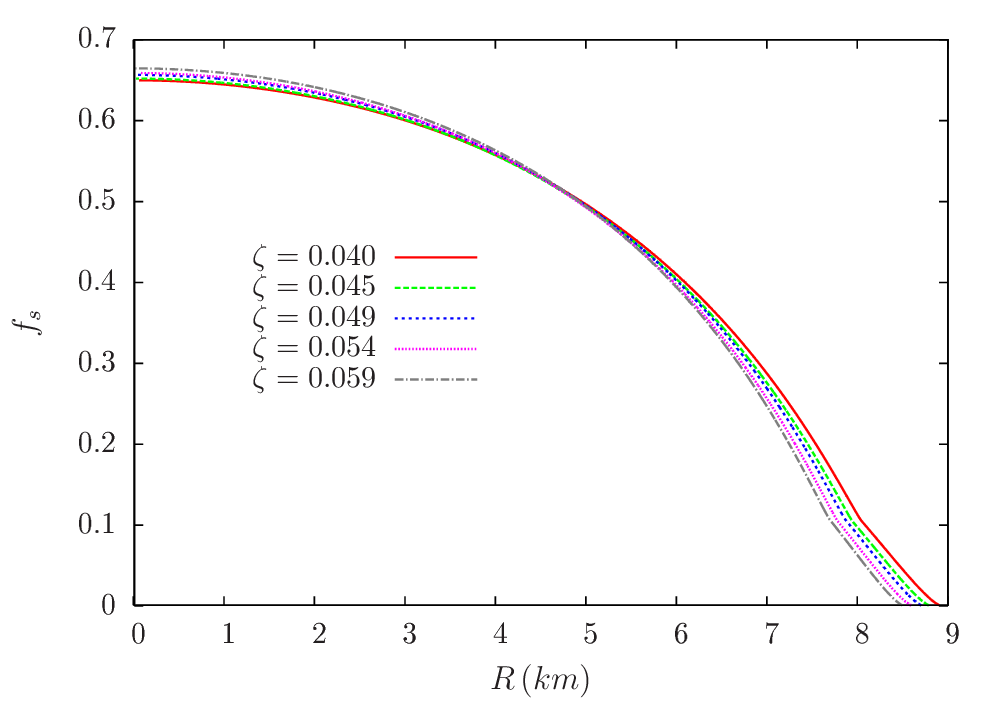}
 \caption{\label{fs_lambda} Fraction of strangeness $f_s$ as a function of the radius for the maximum mass star provided for the $\zeta$-parameter. 
 The hyperon potentials are fixed to $U_{\Lambda}^N=-28 \,\mathrm{MeV}$, $U_{\Sigma}^N=+30\,\mathrm{MeV} $ and $U_{\Xi}^N=-18\,\mathrm{MeV} $, 
 and symmetry energy and its slope at saturation are $a_{sym}^0= 32 \,\mathrm{MeV}$ and $L_0=97\,\mathrm{MeV}$, respectively. 
 We have plotted only the parameterizations that are in agreement with observational data.}
 \end{figure}

In order to specify the strangeness content of the stars, we define the parameter $f_s$:
\begin{equation}\label{strangeness}
f_s= \sum_{i}\frac{\rho_{i}Q_{s_{i}}}{\rho_{b}}=\frac{\rho_{\Lambda}+\rho_{\Sigma}+2\rho_{\Xi}}{\rho_{b}},
\end{equation}
that corresponds simply to the number of strange quarks per baryon in the particle population (where $Q_{s_{i}}$ corresponds to the strangeness charge).
Figure \ref{fs_lambda} shows $f_s$ as a function of 
the radius for the most massive star provided by each parametrization. 
We have plotted the strangeness profiles only for the parameterizations that generate
stars with masses of at least $1.99M_{\odot}$. 
As discussed above, lower values of the $\zeta$ parameter allow for the early appearance
of hyperons in the star (i.e. outer layers). 
On the other hand, the growth of the hyperon content is slower for these cases (small $\zeta$),
providing a slightly lower central fraction of strangeness.
We have computed the strangeness content only for the parameterizations in agreement with 
observational data, from which we conclude that the star's central values of strangeness do not depart much from 
$f_{s,c}=0.65$, for a range of $\zeta=0.040-0.059$. 
Finally, it is important to emphasize that all results concerning the $\zeta$ parameter are directly related to
the effective mass of the nucleon and the compressibility modulus at saturation, 
as already discussed in section III.
 
\subsection{The role of the hyperon potentials on the properties of hyperon stars} \label{Upotentials_section}

Although not much is known about hyperon-hyperon interactions, the hyperon-nucleon interactions
are a little more constrained from hypernuclear data: the $\Lambda N$ interaction has a well-constrained potential
of about $- 28\, \mathrm{MeV}$
 ; the $\Sigma N$ interaction points towards a repulsive potential; the $\Xi N$ interaction
also presents an attractive potential, but not as well constrained as the $\Lambda N$.

Several works have investigated both attractive and repulsive $U_{\Sigma}^N$ potentials 
\cite{Knorren:1995ds,Schaffner:1995th,SchaffnerBielich:2008kb, Mi:2007zz, Negreiros:2010hk,Vasconcellos:2014qua} 
in the description of hadronic matter. In particular, we mention the analysis of the impact of different values of hyperon potentials on
the properties of neutrons stars carried out by Weissenborn et al \cite{Weissenborn:2011kb} and also recently by reference 
\cite{Bhowmick:2014pma}, both using different RMF models. 
Following these works, in this subsection we aim to verify the effects of hyperon potentials in our model.

Initially, we vary $U_{\Lambda}^N$ and $U_{\Sigma}^N$ around $\pm 2\, \mathrm{MeV}$
  and $\pm 20\, \mathrm{MeV}$, respectively.  
Fixing the values of the remaining hyperon potential, we solve the TOV equations for different choices of the $\zeta$ parameter.
From this case, no significant effect on the maximum mass and radius of the neutron stars are found.
The results for our model agree with those found in references \cite{Weissenborn:2011kb,Bhowmick:2014pma} for different RMF models.

\begin{figure}
 \centering
 \vspace{1.0cm}
 \includegraphics[width=9.cm]{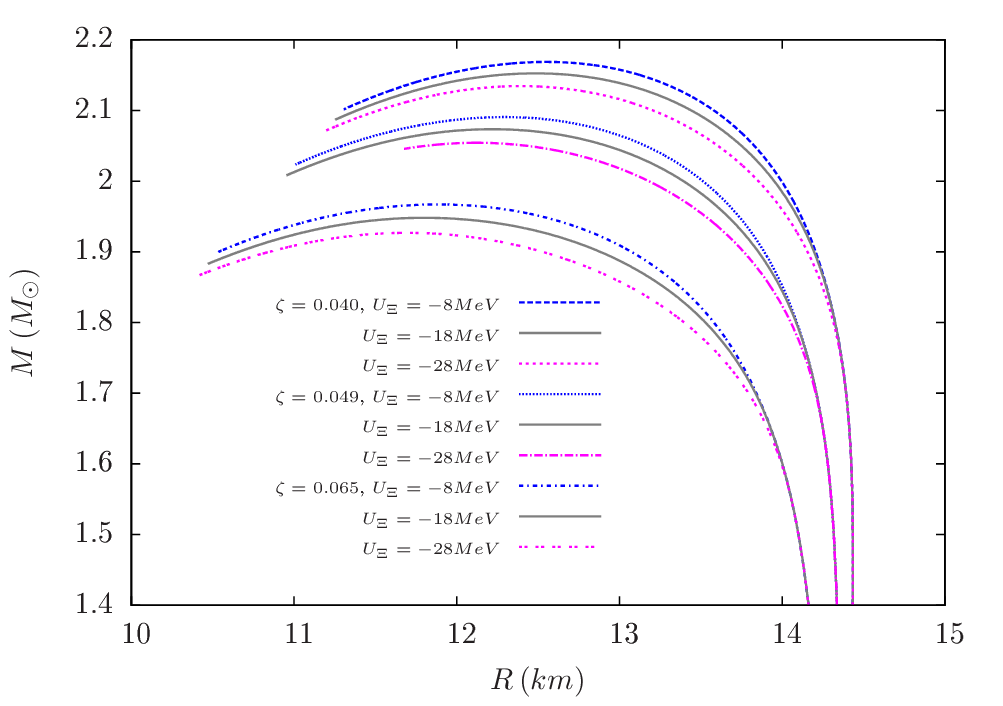}
 \caption{\label{tov_Uc} Mass-Radius relation of hyperonic matter shown for different $U_{\Xi}^N$ potentials and choices of $\zeta$-parameter.}
 \end{figure}

We then varied the $U_{\Xi}^N$ potential around $\pm 10\, \mathrm{MeV}$, from which we find a relevant alteration of the maximum mass
of the stars predicted by the model.
Figure \ref{tov_Uc} shows the Mass-Radius relation for different $U_{\Xi}^N$ potentials and choices of the $\zeta$-parameter. 
The other hyperon potentials are fixed to $U_{\Lambda}^N=-28 \,\mathrm{MeV}$ and $U_{\Sigma}^N=+30\,\mathrm{MeV} $, and the symmetry energy and its slope 
at saturation are $a_{sym}^0= 32\, \mathrm{MeV}$
  and $L_0=97\, \mathrm{MeV}$
 .
The results presented here only include the three choices of the $\zeta$ parameter ($0.040,\,0.049,\,0.065$, corresponding to the
higher, medium and lower branches, respectively) that provide massive stars.
We conclude that, for our model, a change of $10\,\mathrm{MeV}$ in the $U_{\Xi}^N$ potentials results in a change of approximately $0.02M_{\odot}$
in the possible maximum masses of the stars.

\begin{figure}
 \centering
 \vspace{1.0cm}
 \includegraphics[width=9.5cm]{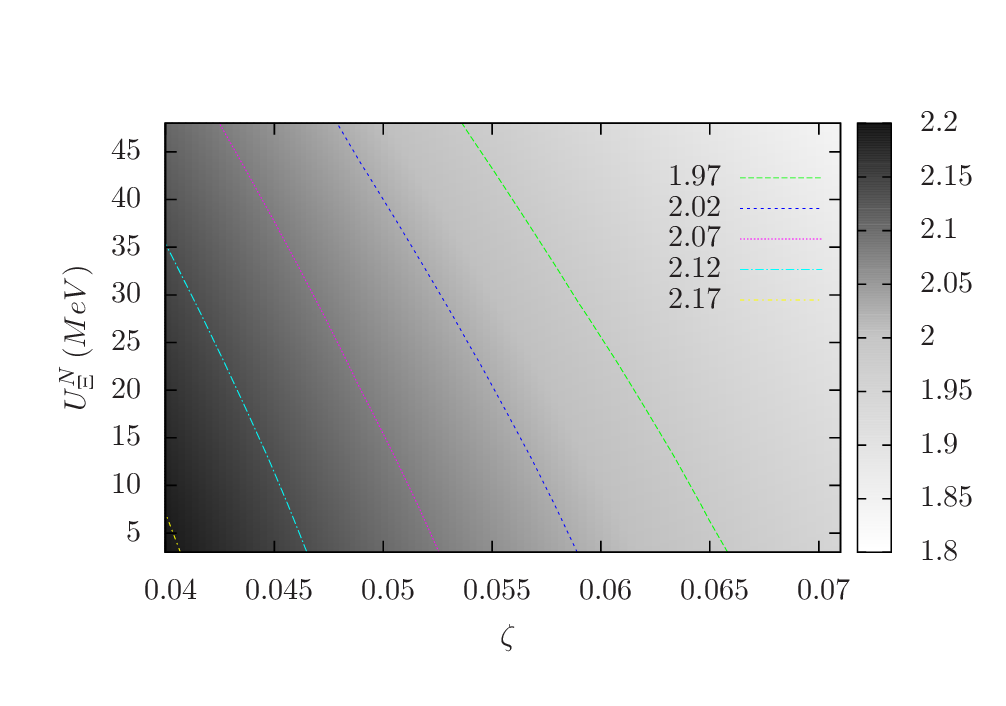}
 \caption{\label{space_Uc_lambda} Parameter space related to the maximum star mass in the model. The horizontal and vertical axes
correspond to the $\zeta$ parameter and the $U_{\Xi}^N$, respectively. 
The other hyperon potentials are fixed to $U_{\Lambda}^N=-28 \,\mathrm{MeV}$ and $U_{\Sigma}^N=+30\,\mathrm{MeV} $, 
 and symmetry energy and its slope at saturation are $a_{sym}^0= 32 \,\mathrm{MeV}$ and $L_0=97\,\mathrm{MeV}$, respectively.}
 \end{figure}

From all of the star properties analyzed so far, we come to the conclusion that the $\zeta$ parameter and the $U_{\Xi}^N$ potential
are the only quantities that have a significant impact on the maximum star mass predicted by 
our model (in the $\sigma \omega \varrho \delta \phi$ version).
For this reason, in Figure \ref{space_Uc_lambda}, we present the parameter space of these quantities related to the 
maximum mass of the stars. The vertical axis corresponds to the $U_{\Xi}^N$ potential in the range of $-3$ to $-48\, \mathrm{MeV}$, 
the horizontal axis corresponds to the $\zeta$ parameter in a range of $0.040-0.071$, and the scale of colors correspond to the stars's maximum
mass (in $M_{\odot}$) provided by each choice of parameters. 
We indicate the limit of $1.97\,M_{\odot}$ that corresponds to the lower mass limit of the pulsar PSR J0348+0432.

The results in Figures \ref{tov_Uc} and \ref{space_Uc_lambda} show that the maximum masses of the stars are higher
as the attraction of the $U_{\Xi}^N$ potential becomes weaker. 
Weak $U_{\Xi}^N$ potentials shift the threshold of the $\Xi$'s to higher densities, leaving hadronic matter to be populated 
mainly by $n p e \mu \Lambda$ degrees of freedom at very high densities. 
In such a scenario, the filling of the energy states turns the EoS very stiff, which ultimately generates 
massive neutron stars.
Note that the results in Figure \ref{space_Uc_lambda}, actually relate nuclear, hypernuclear
and astrophysical data, since the $\zeta$ parameter has a direct relation with the effective mass of the nucleon 
and the compressibility modulus at saturation. 
 
We now turn our attention to the effect of hyperon potentials on the strangeness content inside the stars.
Since the $U_{\Lambda}^N$ potential should not depart much from the value of $- 28\, \mathrm{MeV}$ from experimental data, 
we have verified the particle population of the model changing this value by $\pm 2\, \mathrm{MeV}$.
In this case, no significant change was found.

\begin{figure}
\centering
 \includegraphics[width=9.cm]{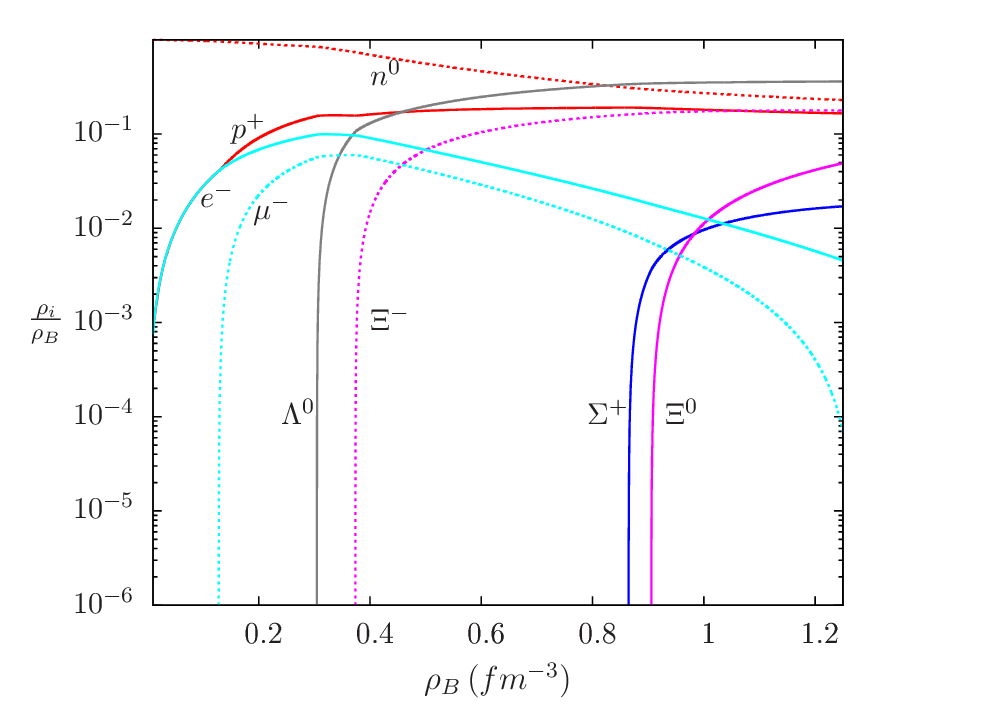}
      \centering
   \includegraphics[width=9.cm]{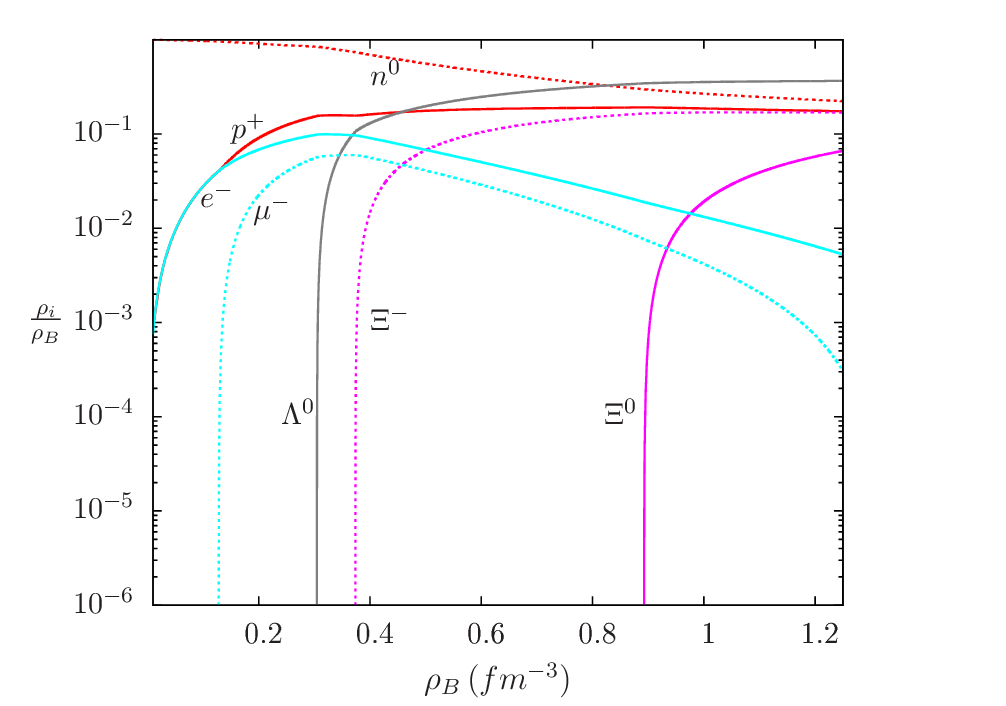}
    \caption{Particle population, for $\zeta=0.040$, dependence on the hyperon potential $U_{\Lambda}$. 
The top and bottom panels correspond respectively to the cases of $U_{\Sigma}^N=+10\, \mathrm{MeV}$
  and $U_{\Sigma}^N=+50\, \mathrm{MeV}$
 .
    The other hyperon potential are fixed to $U_{\Lambda}^N=-28\, \mathrm{MeV}$
  and $U_{\Xi}^N=-18\, \mathrm{MeV}$
 . 
}\label{pop_Us}
 \end{figure}

Figure \ref{pop_Us} shows the particle population for different $U_{\Sigma}^N$ potentials. 
The top panel corresponds to $U_{\Sigma}^N=+10\, \mathrm{MeV}$
  and the bottom one to $U_{\Sigma}^N=+50\, \mathrm{MeV}$, for $\zeta=0.040$.
The results show that the $\Sigma^+$ populate matter only at very high densities ($0.86\,\mathrm{fm}^{-3}$) in the 
case of a weak repulsion of $U_{\Sigma}^N=+10\, \mathrm{MeV}$. 
We report a threshold value of $U_{\Sigma}^N=+26\, \mathrm{MeV}$ for the vanishing of $\Sigma$ particles in our model
in a range of $0-10\rho_0$ (for fixed $U_{\Lambda}^N=-28\, \mathrm{MeV}$ and $U_{\Xi}^N=-18\, \mathrm{MeV}$), which does not depend on the $\zeta$ parameter. 
As $\Sigma$'s appear only at very high densities, that usually exceed 
the central densities in neutron stars in certain models, the value of $U_{\Sigma}^N$ does not bring important astrophysics predictions
concerning neutron stars in those cases, as already pointed out by Schaffner \cite{SchaffnerBielich:2008kb}.
  
\begin{figure}
\centering
\includegraphics[width=9.cm]{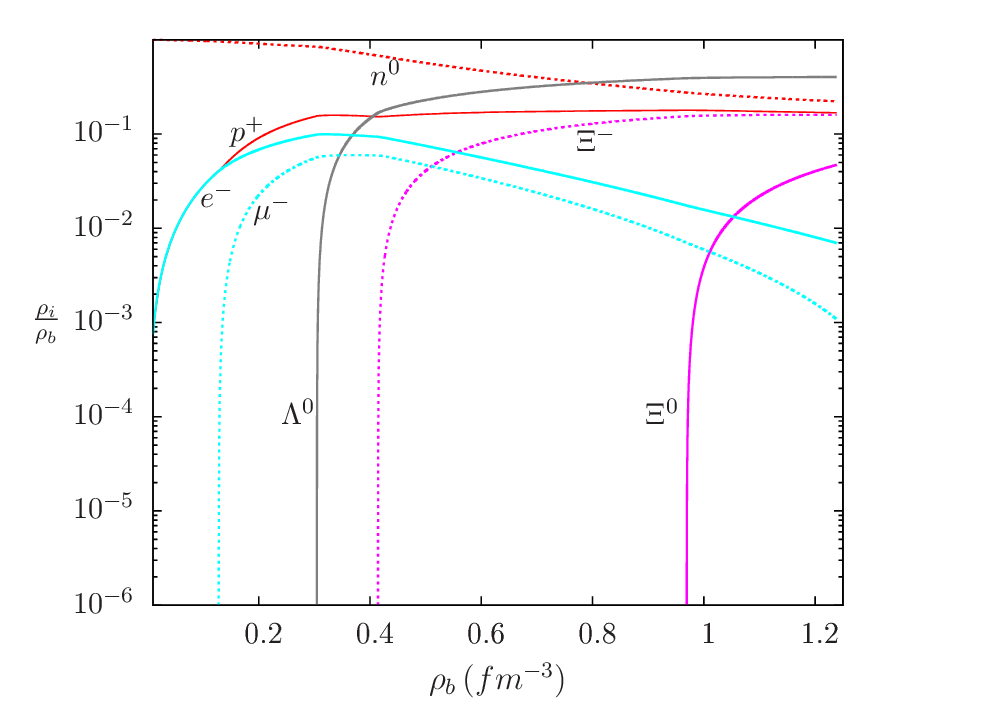}
    \centering
 \includegraphics[width=9.cm]{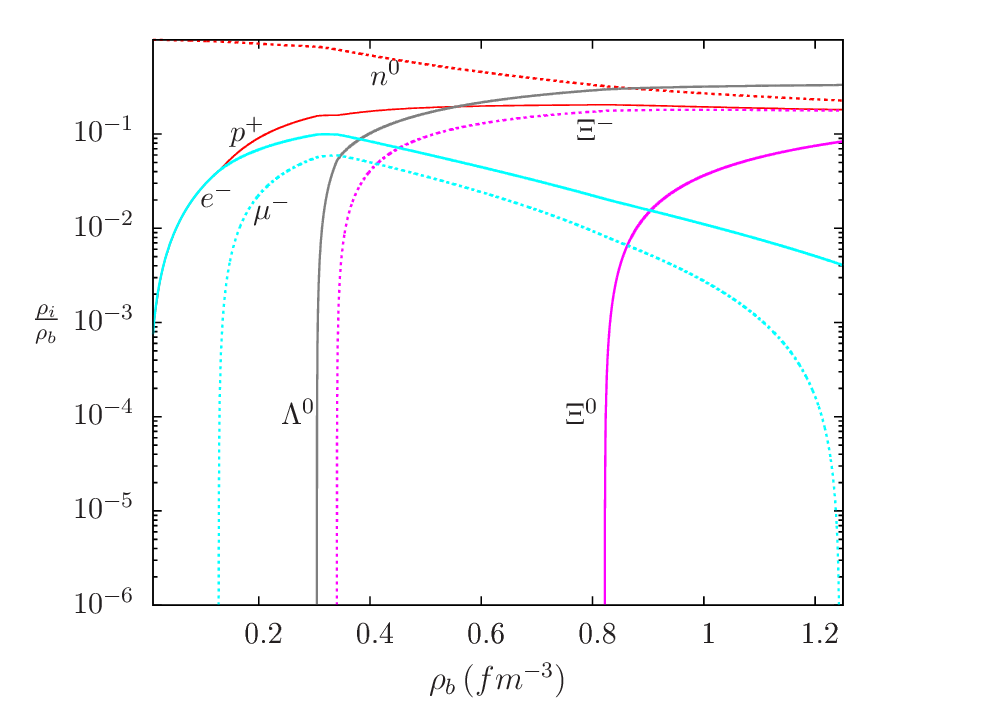}
      \caption{Same as Figure \ref{pop_Us}, but $U_{\Lambda}$, top and bottom panels correspond, respectively, to the cases of $U_{\Xi}^N=-8\, \mathrm{MeV}$
  and $U_{\Xi}^N=-28\, \mathrm{MeV}$
 .
}\label{pop_Uc}
 \end{figure}

The particle population for different $U_{\Xi}^N$ potentials is shown in Figure \ref{pop_Uc}, where 
the top and bottom panels correspond to $U_{\Xi}^N=-8\, \mathrm{MeV}$
  and $U_{\Sigma}^N=-28\, \mathrm{MeV}$, respectively, for $\zeta=0.040$.
It can be seen that an attractive $U_{\Xi}^N$ potential, such as the one at the bottom panel,
pulls the appearance of $\Xi$'s to lower densities. 
Ultimately, a very strong attraction can reverse the order of appearance of the $\Lambda$ and $\Xi^-$ baryons.
In particular, a difference of $20 \,\mathrm{MeV}$ in the $U_{\Xi}^N$ potential shifts the density threshold 
for the appearance of $\Xi^-$ and $\Xi^0$ by about $0.07\,\mathrm{fm}^{-3}$
  and $0.15\,\mathrm{fm}^{-3}$, respectively. 

\begin{figure}
 \centering
 \vspace{1.0cm}
 \includegraphics[width=9.cm]{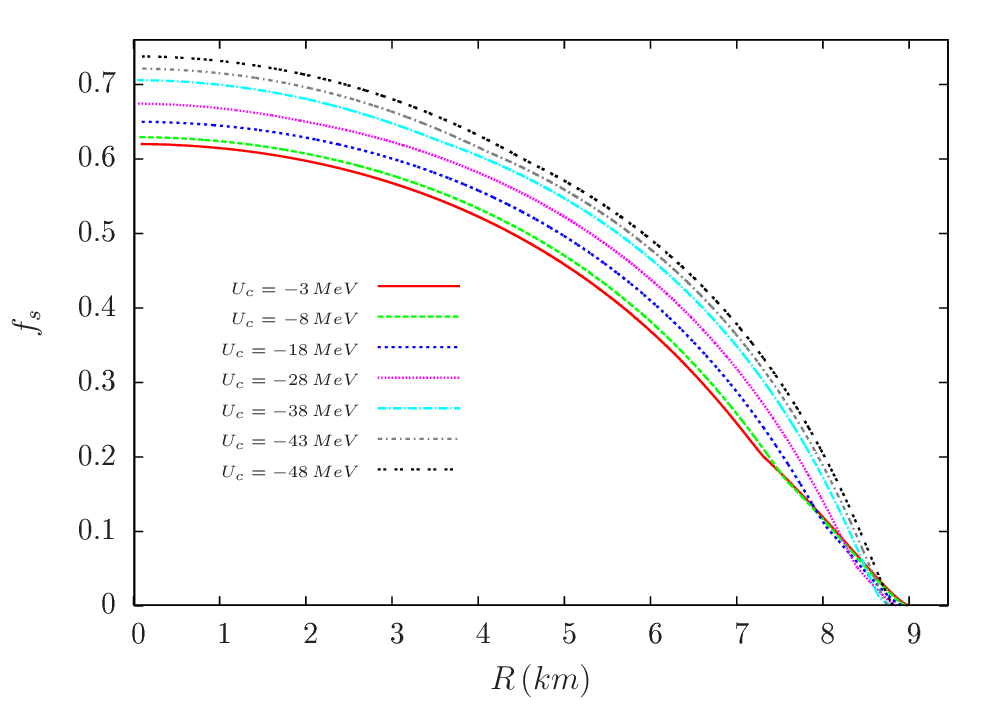}
 \caption{\label{fs_Uc} Fraction of strangeness $f_s$ as a function of radius of the maximum mass star provided by $\zeta=0.040$ 
for different choices of the $U_{\Xi}^N$ potential. 
The other hyperon potentials are fixed to $U_{\Lambda}^N=-28\, \mathrm{MeV}$ and $U_{\Sigma}^N=+30\, \mathrm{MeV} $, 
 and symmetry energy and its slope at saturation are $a_{sym}^0= 32\, \mathrm{MeV}$ and $L_0=97\, \mathrm{MeV}$, respectively.}
 \end{figure}

Since different values of the $U_{\Xi}^N$ potential significantly affect the hyperon population, 
in Figure \ref{fs_Uc} we investigate the strangeness content $f_s$ (vertical axis) as a function of the radius of 
the maximum mass star reproduced (horizontal axis) for fixed $\zeta=0.040$. 
The mass and central densities of the corresponding stars ranges from $2.09M_{\odot}$ and $0.90\,\mathrm{fm}^{-3}$
  (for $U_{\Xi}^N=-48\, \mathrm{MeV}$
 ) 
to $2.18M_{\odot}$ and $0.83\,\mathrm{fm}^{-3}$
  (for $U_{\Xi}^N=-3\, \mathrm{MeV}$
 ). 
The curves in Figure \ref{fs_Uc} indicate an increase in the fraction of strangeness as the attraction of 
the potential $U_{\Xi}^N$ gets stronger. 
Strong attractive $U_{\Xi}^N$ potentials shift the threshold of hyperons appearance to lower densities (higher radius),
leaving a broader range of densities for $f_s$ (hyperon population) to grow. 
Hence, the potentials that allow the appearance of hyperons at lower densities provide 
more strangeness at the center of the star. 
In particular, the range of $-3$ to $-48\, \mathrm{MeV}$
  for $U_{\Xi}^N$ provides central values of $f_s$ of about $0.62-0.74$.
 
Note that, since varying $U_{\Lambda}^N$ and $U_{\Sigma}^N$ generate always the same 
family of stars, the $f_s$ profile is not altered by the change of their values.
Finally, we again stress that the hyperon potentials concerning the hyperon-nucleon (YN) interactions have 
no effect on the radius of neutron stars, in agreement with refs. \cite{Weissenborn:2011kb, Bhowmick:2014pma}. 
This result is not dependent on different choices of parameters of the model.

\subsection{The role of $\sigma^*$-meson on the properties of hyperon stars}  \label{gss_section}
 
In order to introduce the remaining interaction between hyperons due to the meson $\sigma^*$,
we vary the strengh of the coupling constant $g_{\sigma^* Y}$, setting $g_{\sigma^* \Lambda}=g_{\sigma^* \Sigma}=g_{\sigma^* \Xi}$. 
This approach allows us to constrain hyperon-hyperon interactions through astrophysical data investigation, 
differently from fixing a value of the hyperon-hyperon potential ($U_Y^Y$) \cite{Schaffner:1995th,Mi:2010zz}.

\begin{figure}
 \centering
 \vspace{1.0cm}
 \includegraphics[width=9.cm]{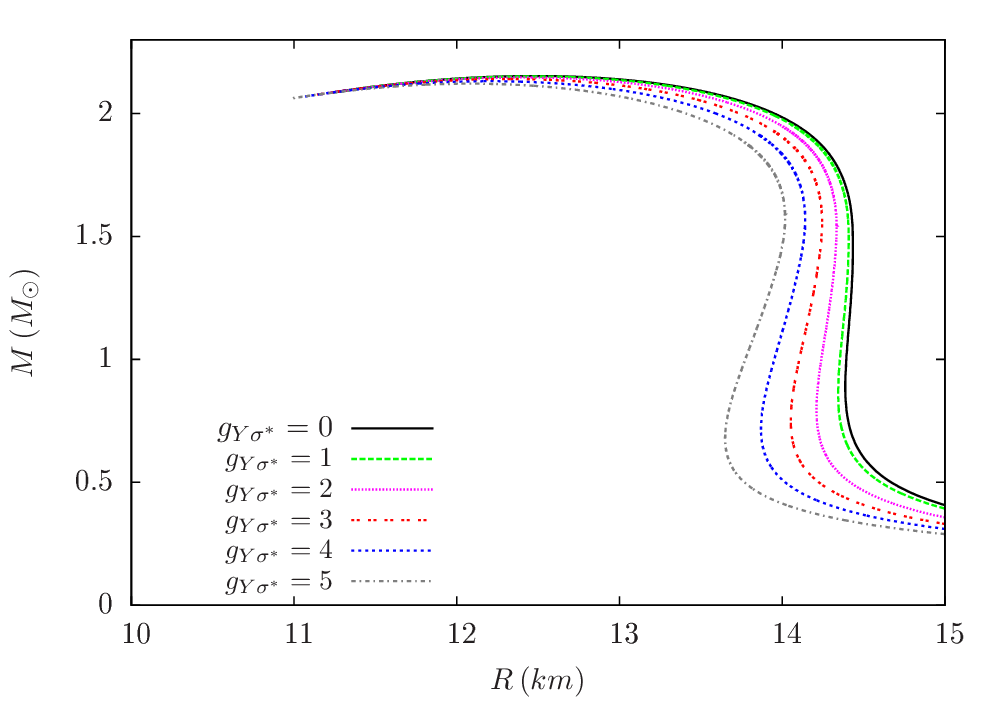}
 \caption{\label{tov_l0040_gss} Mass-Radius relation of hyperonic matter, for $\zeta=0.040$, shown for different choices of $g_{\sigma^* Y}$ coupling constants. 
 The hyperon potentials are fixed to $U_{\Lambda}^N=-28 \,\mathrm{MeV}$, $U_{\Sigma}^N=+30 \,\mathrm{MeV}$ and $U_{\Xi}^N=-18\,\mathrm{MeV} $, and symmetry energy and its slope 
 at saturation are $a_{sym}^0= 32 \,\mathrm{MeV}$ and $L_0=97\,\mathrm{MeV}$, respectively.}
 \end{figure}

To do so, we solve the TOV equations for coupling constants ranging $g_{\sigma^* Y}=0$ to $5.0$, as shown in 
Figure \ref{tov_l0040_gss}, for $\zeta=0.040$.
The results show that the $\sigma^*$ meson has effects on both the maximum mass and the radius of the
stars generated by the model.
The increase of the $\sigma^*$ coupling allows for more attraction in the matter, consequently
lowering the maximum mass and radius predicted by the model. 

\begin{figure}
\centering
\vspace{1.0cm}
 \includegraphics[width=9.cm]{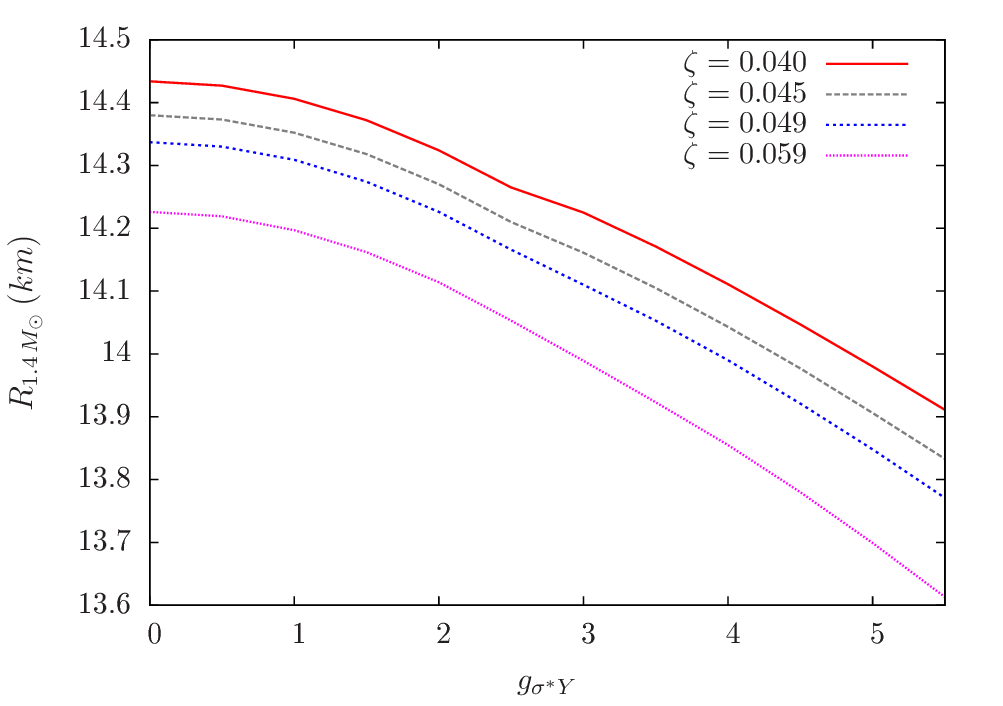}
    \caption{Radius of the $1.4\, M_{\odot}$ star dependence on the intensity of the $g_{\sigma^* Y}$ coupling costants, for different choices of the $\zeta$ parameter. 
 The hyperon potentials are fixed to $U_{\Lambda}^N=-28 \,\mathrm{MeV}$, $U_{\Sigma}^N=+30 \,\mathrm{MeV}$ and $U_{\Xi}^N=-18 \,\mathrm{MeV}$, and symmetry energy and its slope 
 at saturation are $a_{sym}^0= 32 \,\mathrm{MeV}$ and $L_0=97\,\mathrm{MeV}$, respectively.} \label{radius_gss} 
 \end{figure}

In order to better quantify each effect, in Figure \ref{radius_gss} 
we present the change in the radius of the canonical star of $1.4\,M_{\odot}$ as a function 
of the coupling $g_{\sigma^* Y}$. The choices of the parameter $\zeta$ are those which, in the 
absence of the $\sigma^*$ meson, reach the minimum star mass value of $1.97\,M_{\odot}$.
In particular, paying attention to the parameterization that provides the higher neutron star mass ($\zeta=0.040$),
we verify that, for a range $g_{\sigma^* Y}=0-5.5$, the radius of the canonical star decreases by $0.52\,km$ and 
the maximum mass drops from $2.15\,M_{\odot}$ to $2.12\,M_{\odot}$.
It is important to note that the value of the coupling $g_{\sigma^* Y}$ cannot be simply increased, since it 
has a direct impact on the effective mass of the baryons. 
As already pointed out in the literature, at the density in which the effective masses of baryons reaches a zero value, 
it is necessary to use a formalism beyond mean field approximation and purely baryonic matter \cite{Schaffner:1995th,Taurines:2000zb}.

\begin{figure}
 \centering
 \vspace{1.0cm}
 \includegraphics[width=9.5cm]{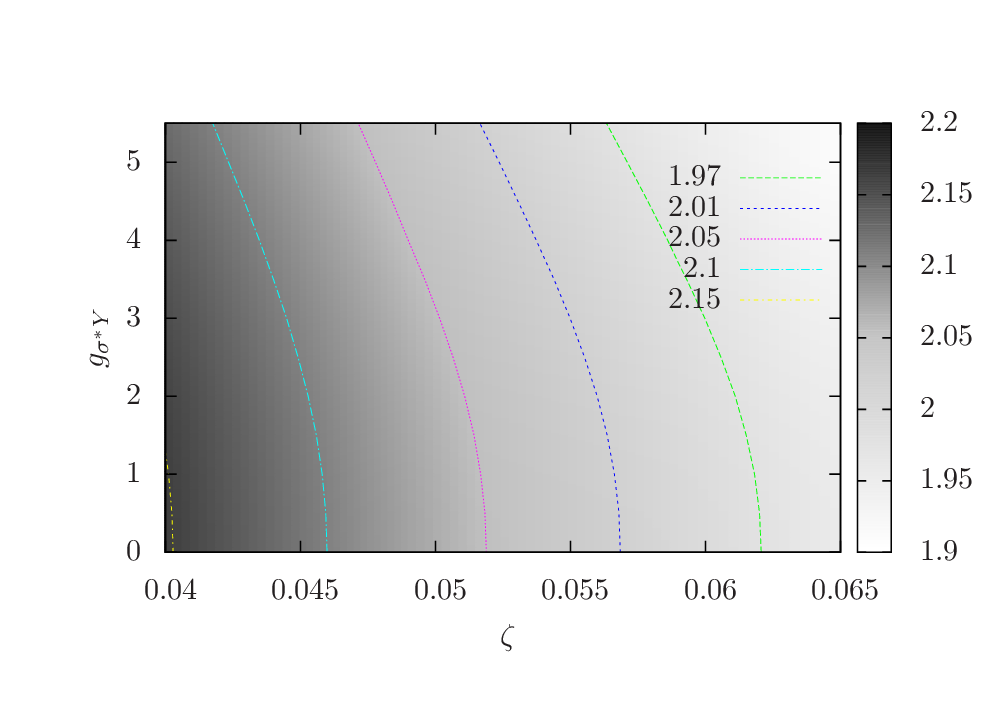}
 \caption{\label{space_gss_lambda} Parameter space related to star's maximum masses in the model. The horizontal and vertical axes
correspond to the $\zeta$ parameter and $g_{\sigma^* Y}$, respectively. 
The hyperon potentials are fixed to $U_{\Lambda}^N=-28 \,\mathrm{MeV}$, $U_{\Sigma}^N=+30 \,\mathrm{MeV}$, $U_{\Xi}^N=-18 \,\mathrm{MeV}$ 
 and symmetry energy and its slope at saturation are $a_{sym}^0= 32 \,\mathrm{MeV}$ and $L_0=97\,\mathrm{MeV}$, respectively.}
 \end{figure}

Analogously to the analysis carried out for the $U_{\Xi}^N$ potential, in Figure \ref{space_gss_lambda}
we present the parameter space concerning the $g_{\sigma^* Y}$ coupling, the $\zeta$ parameter 
(related to $m^*_N$ and $K_0$)
and the maximum mass of stars produced by our model.
As already discussed above, the lower values of $g_{\sigma^* Y}$ and $\zeta$ are those that provide the
highest star masses, as shown in the bottom left region of the plot. 
The threshold of the possible minimum mass is traced by the $1.97\,M_{\odot}$ curve, according to astrophysical data.
  
\begin{figure}
\centering
 \includegraphics[width=9.cm]{pop_paper_l0040_L97a32.png}
      \centering
 \includegraphics[width=9.cm]{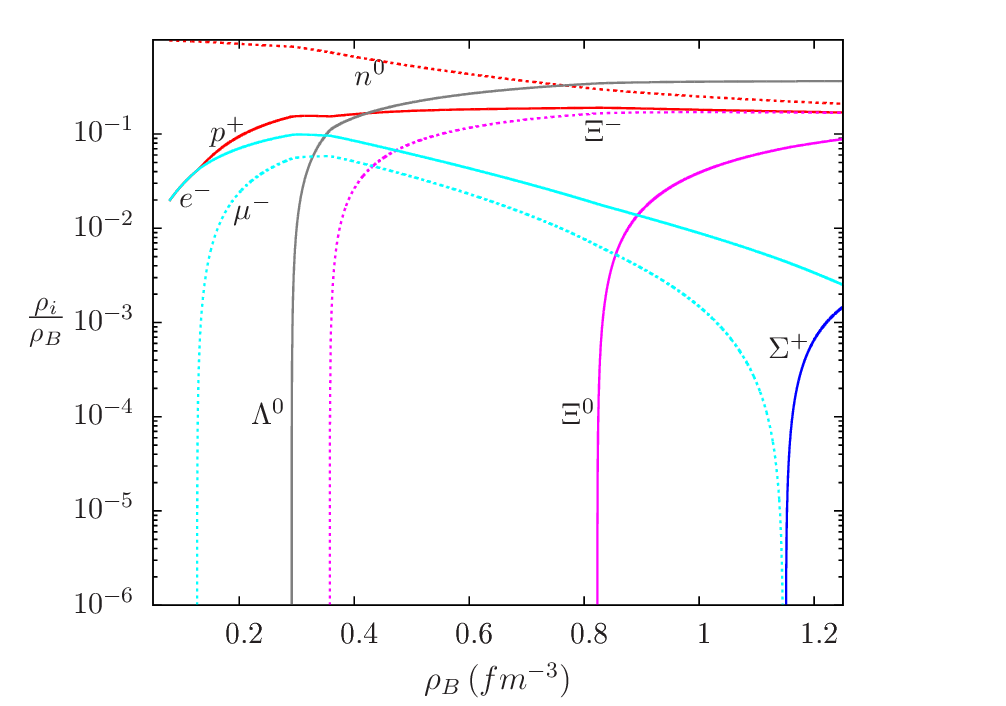}
      \caption{Particle population, for $\zeta=0.040$, with $g_{\sigma^* Y}=0$ (top panel) and $g_{\sigma^* Y}=5$ (bottom panel) 
      coupling strenghts. The hyperon potentials are fixed to $U_{\Lambda}^N=-28\, \mathrm{MeV}$, $U_{\Sigma}^N=+30\, \mathrm{MeV}$
  and $U_{\Xi}^N=-18\, \mathrm{MeV}$ and the symmetry energy and its slope are $a^0_{sym}=32\,\mathrm{MeV}$ and $L_0=97\,\mathrm{MeV}$. } \label{pop_gss}
\end{figure}

We also analyze the effect of the $g_{\sigma^* Y}$ coupling on the particle population in 
Figure \ref{pop_gss}. The top and bottom panels show the population in 
the absence of the $\sigma^*$ meson and the case in which $g_{\sigma^* Y}=5.0$ (for $\zeta=0.040$).
The $g_{\sigma^* Y}$ coupling has impact only on the particle population at high densities, 
shifting the threshold of appearance of $\Xi^0$ to lower densities (from about $0.9\,\mathrm{fm}^{-3}$
  to $0.82\,\mathrm{fm}^{-3}$
 ).

More dramatically, the extra attraction introduced by the $\sigma^*$ meson pushes the $\Sigma^+$ hyperon threshold 
back to the range of densities between $0-10\rho_0$. In particular, for $g_{\sigma^* Y}=5.0$, these
particles appear at $1.15\,\mathrm{fm}^{-3}$, which cannot quite be reached in neutron stars.
Moreover, as such high densities are only reached in the core of neutron stars, we also verified
that the $g_{\sigma^* Y}$ coupling does not have significant effects on the strangeness profile of the stars.

Finally, we show in Table \ref{tab_Ypotentials} the change in the values of the hyperon-hyperon potetials due to the 
introduction of the $\sigma^*$ meson with a coupling $g_{\sigma^* Y}$.
In the second column we show the nucleon-nucleon potential $U^N_N$ that becomes less attractive for higher values of the $\zeta$ parameter,
changing from $-66.90\, \mathrm{MeV}$
  for $\zeta=0.040$ to $-64.47\, \mathrm{MeV}$
  for $\zeta=0.059$. 
The value of $U_N^N$ continues decreasing for higher values of $\zeta$, reaching $-59.56\, \mathrm{MeV}$
  for $\zeta=0.129$, 
which corresponds to the limiting value to reproduce the properties of nuclear matter at saturation density. 

The introduction of the $\sigma^*$ meson produces a stronger attraction among the $\Lambda$ particles, 
changing $U_{\Lambda}^{\Lambda}$ from $+ 4.54\, \mathrm{MeV}$
  to $-20.86\, \mathrm{MeV}$
  for $\zeta$ and from $-2.58\, \mathrm{MeV}$
  to $-27.23\, \mathrm{MeV}$
  for $\zeta=0.059$.
Like the nucleon-nucleon potential, the value of $U_{\Lambda}^{\Lambda}$ continues to decrease for higher values
of $\zeta$, reaching $-14.86\, \mathrm{MeV}$
  (for $g_{\sigma^* Y}=0$) and $-38.19\, \mathrm{MeV}$
  (for $g_{\sigma^* Y}=5$), for $\zeta=0.129$.
All the other hyperons present repulsive interactions, even with the extra attraction introduced by the $\sigma^*$ meson.
Although the only measurement of the double-$\Lambda$ hypernuclei points towards a weak  $\Lambda \Lambda$ depth potential \cite{Millener:1988hp},
only more hypernuclear data will be able to better constrain the $\sigma^*$ coupling constants.
The same applies to the other $YY$ potentials, from which, so far, no accurate experimental data exist.

  \begin{table*}[t]
  \centering
  \scriptsize
  \begin{tabular}{*{11}{c}}
\hline  \hline
&  &  &  &  &  &  &  &  &  &   \\

\bfseries  hyperon potentials/ & $U^{N}_{N}\,(\mathrm{MeV})$ & $U^{\Lambda}_{\Lambda}\,(\mathrm{MeV})$ & $U^{\Lambda}_{\Sigma}\,(\mathrm{MeV})$& $U^{\Lambda}_{\Xi}\,(\mathrm{MeV})$
& $U^{\Sigma}_{\Lambda}\,(\mathrm{MeV})$ & $U^{\Sigma}_{\Sigma}\,(\mathrm{MeV})$& $U^{\Sigma}_{\Xi}\,(\mathrm{MeV})$& $U^{\Xi}_{\Lambda}\,(\mathrm{MeV})$ 
& $U^{\Xi}_{\Sigma}\,(\mathrm{MeV})$& $U^{\Xi}_{\Xi}\,(\mathrm{MeV})$\\
\bfseries  $\zeta$  &  &  &  &  &  &  &  &  &  &   \\
        \hline \hline 
    &  &  &  &  &  &  &  &  &  &   \\
$\zeta=0.040$, $g_{\sigma^* Y}=0$  & $-66.90$ & $+4.54$ & $+46.25$ & $+49.89$ & $+41.54$ & $+72.48$ & $+69.13$ & $+44.72$ & $+67.92$ & $+121.75$  \\
$\zeta=0.040$, $g_{\sigma^* Y}=5$  & $-66.90$ & $-20.86$ & $+20.53$ & $+23.96$ & $+15.10$ & $+45.70$ & $+42.12$ & $+17.87$ & $+40.70$ & $+95$  \\
&  &  &  &  &  &  &  &  &  &   \\
\hline
&  &  &  &  &  &  &  &  &  &   \\
$\zeta=0.049$, $g_{\sigma^* Y}=0$  & $-65.79$ & $+0.95$ & $+43.25$ & $+44.58$ & $+38.27$ & $+69.02$ & $+64.03$ & $+39.14$ & $+62.87$ & $+111.98$  \\
$\zeta=0.049$, $g_{\sigma^* Y}=5$  & $-65.79$ & $-24.06$ & $+17.85$ & $+18.98$ & $+12.13$ & $+42.46$ & $+37.24$ & $+12.53$ & $+35.83$ & $+84.71$  \\
&  &  &  &  &  &  &  &  &  &   \\
\hline
&  &  &  &  &  &  &  &  &  &   \\
$\zeta=0.059$, $g_{\sigma^* Y}=0$  & $-64.47$ & $-2.58$ & $+40.36$ & $+39.34$ & $+35.14$ & $+65.65$ & $+59.06$ & $+33.71$ & $+58.01$ & $+102.36$  \\
$\zeta=0.059$, $g_{\sigma^* Y}=5$  & $-64.47$ & $-27.23$ & $+15.26$ & $+14.03$ & $+9.25$ & $+39.25$ & $+32.44$ & $+7.29$ & $+31.10$ & $+72.53$  \\
&  &  &  &  &  &  &  &  &  &   \\
\hline \hline
  \end{tabular}
\caption{ Hyperon-hyperon potentials for different values of the $\zeta$ parameter and $g_{\sigma^* Y}$. 
The first column corresponds to the chosen parameterizations.
In the second column we show the nucleon-nucleon potential refering to each parameterization.
The third, fourth and fifth columns show the corresponding values of the $\Lambda$, $\Sigma$ and $\Xi$ potentials with respect to the $\Lambda$ matter.
The sixth, seventh and eighth columns show the corresponding values of the $\Lambda$, $\Sigma$ and $\Xi$ potentials with respect to the $\Sigma$ matter.
The ninth, tenth and eleventh columns show the corresponding values of the $\Lambda$, $\Sigma$ and $\Xi$ potentials with respect to the $\Xi$ matter.
The nucleon-hyperon potentials and the symmetry energy and its slope at saturation are fixed to
$U_{\Lambda}^N=-28\, \mathrm{MeV}$, $U_{\Sigma}^N=+30\, \mathrm{MeV}$ and $U_{\Xi}^N=-18\, \mathrm{MeV}$, $a_{sym}^0= 32\, \mathrm{MeV}$ and $L_0=97\, \mathrm{MeV}$. } \label{tab_Ypotentials}
\end{table*}
\normalsize

\subsection{The radius of the canonical star} \label{radius_section}

Thus far, in this paper, we have investigated the effects of the parameters of the formalism on
the properties of neutron stars, except the radius.
Since most measured neutron star masses are clustered around 
$1.4M_{\odot}$, we now study all quantities that have impact on the radius of these stars.

We have discussed in previous sections that the parameter $\zeta$, which relates the effective mass
of the nucleon to the compressibility modulus at saturation, alters both the mass and the radius of compact stars.
As already explained, $\zeta$ reflects how the many-body forces influence nuclear matter.
In Section \ref{prop_high_section} we have shown that this parameter
has an impact on the behavior of star matter properties at high densities,
such as the density dependence of the compressibility modulus, symmetry energy and its slope.

Lopes et al \cite{Lopes:2014wda} showed how an arbitrary variation of the slope $L_0$ changes the radius of the
canonical star. In the particular case of our model, we verify that the only quantities that affect the behavior of
the symmetry energy $a_{sym}$ and its slope $L$ at high densities are the $\zeta$ parameter and 
the constraints of these values ($a_{sym}^0$ and $L_0$) at saturation. 
From this, we conclude that in our case the variation of the slope $L_0$ with respect to different densities, 
would arise from different behavior of many-body forces, for fixed values of $a_{sym}^0$ and $L_0$.

\begin{figure}
 \centering
 \vspace{1.0cm}
 \includegraphics[width=9.5cm]{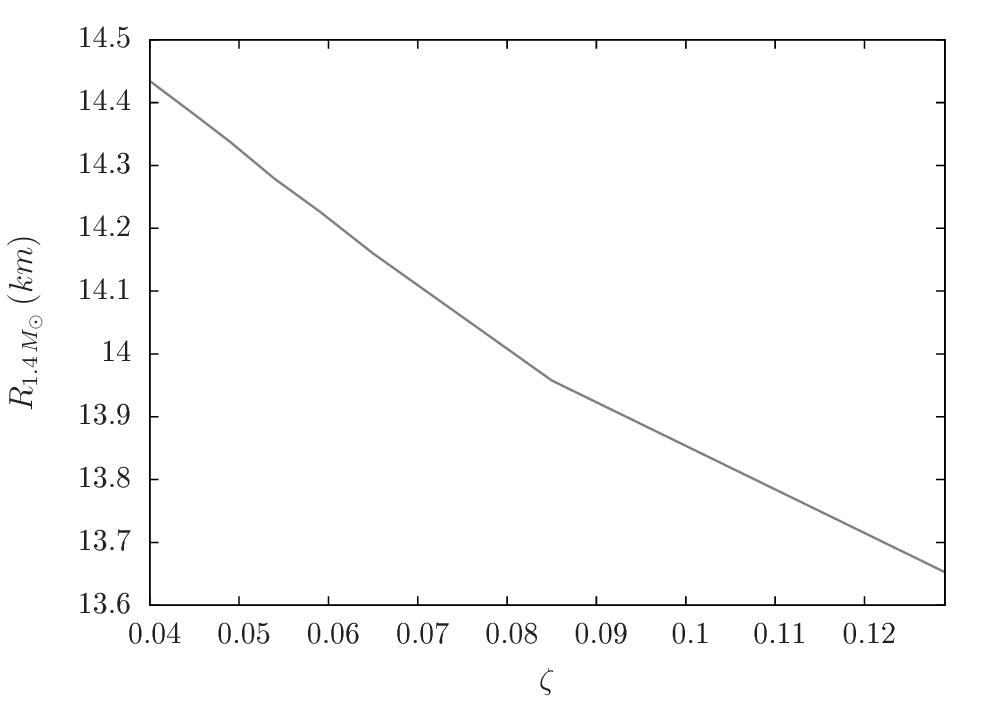}
 \caption{\label{radius14_lambda} Effect on the $\zeta$ parameter on the radius of the canonical star ($1.4\,M_{\odot}$). 
 The horizontal and vertical axes refer to the $\zeta$ parameter and the radius of the star, respectively. 
The hyperon potentials are fixed to $U_{\Lambda}^N=-28 \,\mathrm{MeV}$, $U_{\Sigma}^N=+30 \,\mathrm{MeV}$, $U_{\Xi}^N=-18 \,\mathrm{MeV}$ 
 and symmetry energy and its slope at saturation are $a_{sym}^0= 32 \,\mathrm{MeV}$ and $L_0=97\,\mathrm{MeV}$, respectively.}
 \end{figure}

Figure \ref{radius14_lambda} shows the change of the radius of the canonical star as a function of the 
$\zeta$ parameter. Again, we stress that the change of the radius is directly related to the values
of the effective mass and the compressibility modulus at saturation in this model.
We observe that, for the range of values that fits nuclear data, the radius of the star changes 
about $0.8\,km$, but when we impose the observational limit of $1.97\,M_{\odot}$ (corresponding to the parameterization $\zeta=0.04-0.059$), 
this value drops to $0.2\,km$. 

 Since the values of the symmetry energy and its slope also alter the radius of the stars, 
 in Table \ref{tab_radius} we quantify changes on the radius considering uncertainties in these values.
 Also, as the $\zeta$ parameter was shown to also affect the radius of neutron stars, we show these
 changes on the radius for the three parameterizations $\zeta=0.040,\,0.049,\,0.059$.
 
 For a $1.4M_{\odot}$ star that belongs to the family that predicts at least $1.97M_{\odot}$ maximum mass stars, 
 we find that the smallest radius is $13.725\,km$, for the parameterization $\zeta=0.059$, $a_{sym}=32\,\mathrm{MeV}$,
 $L_0=94\,\mathrm{MeV}$ and $g_{\sigma^* Y}=4.0$.
 Relaxing the values of the asymmetry properties at saturation, the smallest radius of the  $1.4M_{\odot}$ star
 drops to $13.54\,km$, for the parameterization $\zeta=0.059$, $a_{sym}=29\,\mathrm{MeV}$,
 $L_0=85\,\mathrm{MeV}$ and $g_{\sigma^* Y}=5.2$.
 Note that, although smaller values of $a_{sym}$ increase the radius of the stars for fixed values of slope $L_0$, 
 these smaller values of $a_{sym}$ also allow for smaller values of $L_0$ (as already shown in Figures \ref{map_l0040}, \ref{map_l0071} and \ref{map_l0129})
 and, thus, smaller radii.
 

 \begin{table*}[t]
  \centering
  \begin{tabular}{*{9}{c}}
    \hline  \hline
\bfseries  $\Delta R_{1.4M_{\odot}}$ & & &  &  $\zeta=0.040$&  &  $\zeta=0.049$&  &  $\zeta=0.059$ \\
    \hline 
    & & & & & & & &\\
$[R_{1.4M_{\odot}}(L_0=110\,\mathrm{MeV})-R_{1.4M_{\odot}}(L_0=97\,\mathrm{MeV})]$ &  &  &  &  $0.33$ km  &  & $0.35$ km  &  &  $0.37$ km    \\                    
$[R_{1.4M_{\odot}}(a_{sym}^0=30\,\mathrm{MeV})-R_{1.4M_{\odot}}(a_{sym}^0=33\,\mathrm{MeV})]$ &  &  &  &  $0.12$ km  &  &  $0.13$ km  &  &  $0.15$ km \\
$[R_{1.4M_{\odot}}(g_{\sigma^* Y}=0)-R_{1.4M_{\odot}}(g_{\sigma^* Y}=5.5)]$ &  &  &  &  $0.52$ km  &   & $0.57$ km  &  &  $0.62$ km \\  
    \hline \hline
  \end{tabular}
\caption{ Summary of the properties that affect star radii in the model. The results are shown for hyperon stars
for different parameterizations. The hyperon potentials are fixed to $U_{\Lambda}^N=-28 \,\mathrm{MeV}$, $U_{\Sigma}^N=+30 \,\mathrm{MeV}$, $U_{\Xi}^N=-18 \,\mathrm{MeV}$ 
and the symmetry energy, the slope of the symmetry energy and the hyperon-hyperon coupling constant in relation to the $\sigma^*$ meson are varied.
In the first line, the symmetry energy  is fixed to $a_{sym}^0=32 \,\mathrm{MeV}$ and $g_{\sigma^* Y}=0$. 
In the second line, the slope of the symmetry energy is fixed to $L_{0}=101 \,\mathrm{MeV}$ and $g_{\sigma^* Y}=0$.
In the third line, the symmetry energy and its slope are fixed to $a_{sym}^0=32 \,\mathrm{MeV}$ and $L_{0}=97 \,\mathrm{MeV}$. 
The results are shown only for the parameterizations in agreement with the observational data.} \label{tab_radius}
\end{table*}


\section{Summary and Conclusions}

Substantial efforts have been made to determine the behavior of nuclear matter at high densities.
Recent observations of massive neutron stars 
have renewed the interest in such studies. 
Given these massive stars, it has been suggested that hyperonic matter 
might not exist in compact stars, due to a possible excessive softnening of the EoS caused by these new degrees of freedom. 
Moreover, the comparison between the radius of nucleonic and hyperon stars predicted by several models
has been reviewed, indicating that the radius of hyperon stars is substantialy larger than
models fitted for nucleonic matter only \cite{Fortin:2014mya}.
In this work we extended the formalism proposed by Taurines et al \cite{Taurines:2000zb} 
in order to describe hyperon stars. 
In the light of this extended model, we succeeded in describing neutron stars
that agree with recent observations, despite the fact that it contains a considerable 
amount of hyperonic matter.

We have developed a new class of EoSs that allows for the presence of hyperons, and in which 
the baryon interactions are mediated by mesonic fields in a parametric derivative coupling. 
We extended the original version of the model by considering the complete set of 
scalar-isoscalar ($\sigma$,$\,\sigma^*$), vector-isoscalar ($\omega$,$\,\phi$), vector-isovector ($\varrho$) 
and scalar-isovector ($\delta$) meson fields.
We introduced the $\delta$, $\sigma^*$ and $\phi$ mesons, since the first allows for a better extrapolation
to asymmetric matter and the last two mesons play an important role in the description of hyperon interactions.

This approach allowed us to take nuclear medium effects into account, as the derivative coupling introduces
an analogy to many-body forces, characterized by a single ($\zeta$) parameter.
The many-body contributions were introduced as nonlinear terms contributions to the \emph{effective coupling constants} 
of the model, whose effect is to turn them indirectly density dependent and also to lower the effective masses of the baryons.
We also point out that, since the density dependence of the couplings comes from the scalar fields in  
this formalism, we avoid the rearrangement terms necessary in explicit density dependent formalisms \cite{Typel:1999yq}.

The decrease of the coupling constants and the effective mass of the nucleon as a function of density 
has interesting phenomenological consequences, as it relates to the restoration of chiral symmetry and asymptotic freedom.
However, an extensive analysis concerning this behavior must be carried out in detail in a future publication.

Each parameterization of the model generates a new equation of state and, for particular parameterizations, 
it is possible to describe models already present in the literature such as \cite{Walecka1986, Zimanyi:1990np}.
Initially, we determined the connection between the $\zeta$ parameter and symmetric nuclear matter properties at saturation,
from which we concluded that smaller values of the $\zeta$ parameter allows for lower (higher) nucleonic effective masses (compressibility modulus).
In particular, we pointed out that the parameter $\zeta$ allows to determine both the effective mass of the nucleon and the  
compressibility modulus, differently from other models that need an extra parameter to fix these values \cite{Boguta:1977xi,Typel:1999yq}.

We have also analysed the parameter space that relates the symmetry energy $a_{sym}^0$ and its slope $L_0$ to 
the coupling constants $g_{\varrho N}$ and $g_{\delta N}$ in order to determine the coupling constants of the isovector mesons to the 
nucleon at saturation density. 
Finally, we calculated the volume part of the isospin incompressibility and the skewneess of the 
symmetry energy, from which we concluded that several parameterizations of the  model are in good agreement with the tests 
carried out in the literature \cite{Dutra:2014qga}.

Choosing the parameters according to nuclear matter saturation properties and the available hypernuclear data, 
we concluded that smaller values of $\zeta$ yield stiffer EoS's. 
We verified that different parameterizations also yield quantitatively different particle populations,
but in all cases the hyperon population threshold density was kept at about $\sim 2 \rho_0$.

In order to further validate the microscopic model used, we compared macroscopic predictions with observed data. 
We calculated the Mass-Radius diagram for the parameterizations able to describe nuclear saturation properties. 
We have found that only the parameterizations with $0.040 \leq \zeta \leq 0.059$ were able to match recently 
observed masses of objects PSR J038+0432 ($M = 2.01 \pm 0.04 M_\odot$, \cite{Antoniadis2013}) 
and PSR J1614-2230 ($M = 1.97 \pm 0.04 M_\odot$,\cite{Demorest2010}), for fixed values of hyperon potentials.  
We have demonstrated that, since the value of the coupling constant $g_{\delta N}$ must remain small in order to ensure
lower values of the slope $L_0$, the introduction of the $\delta$ meson does not have a strong effect on the maximum mass of the stars.
On the other hand, we showed that the inclusion of the $\phi$ meson is crucial for the description of a
$2M_{\odot}$ hyperon star.

Following previous works \cite{Weissenborn:2011kb,Bhowmick:2014pma}, we have calculated the dependence of the star's observational properties
on the hyperon potentials. Our results support those found in the literature, in which only the $U_{\Xi}^{N}$
potential has a significant effect on the maximum mass of stars and, none of the hyperon potentials affect the radius of the stars.
In order to find all quantities that modify star masses, we generated the 
parameter space that relates $U_{\Xi}^{N}$, $\zeta$ and $M_{max}$, or hypernuclear, nuclear and astrophysical observational data.

We carried out a similar analysis concerning the $\sigma^*$ meson, from which a new parameter space, 
relating $g_{\sigma^* Y}$, $\zeta$ and $M_{max}$, was generated.
As far as we know, we report for the first time that nonzero values of the $g_{\sigma^* Y}$ coupling
decrease the radius of neutron stars significantly. 
We summarized the effects of all properties present in our model that modify the radius of the canonical star 
and we concluded that the many-body forces parameter contributes to the behavior of the nuclear asymmetric properties at high densities,
which are reflected in the radii of these stars.

We must still make a final remark regarding the limitations of the formalism that we have developed and, 
most importantly, its uncertainties regarding the description of hyperon stars.
First, we developed a model to describe nuclear matter at high densities by extrapolating the
behavior of symmetric matter at saturation density to highly asymmetric matter at densities 
of about $8-10$ times saturation density \cite{Schaffner:1995th}.
Also, the introduction of hyperons in the system brings uncertainties related to the 
poor data from hypernuclear matter, such as the hyperon potentials, the assumption of a $SU(6)$ 
symmetry and the YY interacion concerning the $\sigma^*$\cite{Fortin:2014mya}. 
There are works in the literature that consider approaches beyond $SU(6)$ \cite{Mi:2007zz,Weissenborn:2011ut,Lopes:2013cpa}
and the universality of $g_{\sigma^* \Lambda}=g_{\sigma^* \Xi}=g_{\sigma^* \Sigma}$ \cite{SchaffnerBielich:2000wj,Mi:2010zz,Gusakov:2014ota}.
However, only new data will allow for a better understanding of hyperon matter at very high densities, 
for instance, from the possible measurements of multi-hyper nuclei such as those that will be provided by FAIR in the near future,
and also from the analysis of hyperon-hyperon correlation in heavy-ion collisions (as originally proposed in \cite{Greiner:1989ig}),
and improved lattice QCD calculations of the hyperon-hyperon potentials \cite{Inoue:2010hs}.
Also, similarly, only new accurate observational data may provide
reliable information regarding the radii of compact stars, which are extremely 
important for contraining the EoS of nuclear matter at high densities.

Our purpose with this study was to develop a new model for nuclear matter that takes into account nonlinear terms that
simulate many-body forces and apply the formalism to describe hyperon stars in accordance with recent observations \cite{Antoniadis2013,Demorest2010}. 
A very straightforward extension of this work is the inclusion of nonlinear contributions to the coupling of isoscalar-vector 
and isovector-vector mesons, which requires a new analysis of saturation and asymmetric matter properties in the model.
Also, in a future work, we plan to investigate the thermal evolution of such stars, which depends on their particle composition.
Furthermore, we plan to investigate the phase transition to quark matter, which may take place in the core of high density neutron stars, 
and the effects of magnetic field on the microscopic and macroscopic properties of such stars.
This work is already in progress.

\section{Acknowledgements}
R.O. Gomes and C.A.Z. Vasconcellos would like to thank J. E. Horvath for the suggestion of the parameter space analysis. 
This work is partially supported by grant Nr. BEX 14116/13-8 of the PDSE CAPES and Science without Borders programs 
which are an initiative of the Brazilian Government.
V. Dexheimer would like to acknowledge the support of HIC for FAIR and thank D. Menezes for her suggestions.

\bibliography{deltans.bib}

\end{document}